\documentclass[conference]{IEEEtran}

\usepackage{amsthm, amsmath}
\usepackage{float}
\usepackage{array}
\usepackage{booktabs}
\usepackage{graphicx}
\usepackage{subcaption}
\usepackage[justification=centering,font=small,labelfont=bf]{caption}
\usepackage{subcaption}
\usepackage{indentfirst}
\usepackage{hhline}
\usepackage{fancyhdr}
\usepackage[export]{adjustbox}
\usepackage{tabularx}
\usepackage[utf8]{inputenc}
\usepackage{verbatim} 

\usepackage{multirow}
\newcommand{\PreserveBackslash}[1]{\let\temp=\\#1\let\\=\temp}
\newcolumntype{L}[1]{>{\raggedright\let\newline\\\arraybackslash\hspace{0pt}}m{#1}}
\newcolumntype{C}[1]{>{\centering\let\newline\\\arraybackslash\hspace{0pt}}m{#1}}
\newcolumntype{R}[1]{>{\raggedleft\let\newline\\\arraybackslash\hspace{0pt}}m{#1}}
\usepackage[para,flushleft]{threeparttable}
\usepackage{longtable}
\usepackage{tablefootnote}

\usepackage{array}
\usepackage{url}

\usepackage{balance}

\usepackage{xcolor,colortbl}
\definecolor{codegreen}{rgb}{0,0.6,0}
\definecolor{codegray}{rgb}{0.5,0.5,0.5}
\definecolor{codepurple}{rgb}{0.58,0,0.82}
\definecolor{backcolour}{rgb}{0.95,0.95,0.92}
\definecolor{LightBlue}{rgb}{0.83, 0.91, 1}
\definecolor{LightGreen}{rgb}{0.8, 1, 0.8}
\definecolor{LightPink}{rgb}{1, 0.8, 0.88}
\definecolor{LightYellow}{rgb}{1, 1, 0.6}
\definecolor{light-gray}{gray}{0.9}

\newcommand{\wname}{CoMeFa}
\newcommand{\wnameRAM}{CoMeFa RAM}
\newcommand{\wnameRAMs}{CoMeFa RAMs}

\newcommand{\wnameD}{CoMeFa-D}
\newcommand{\wnameA}{CoMeFa-A}

\usepackage{listings}
\lstdefinestyle{mystyle}{
    backgroundcolor=\color{backcolour},   
    commentstyle=\color{codegreen},
    keywordstyle=\color{magenta},
    numberstyle=\tiny\color{codegray},
    stringstyle=\color{codepurple},
    basicstyle=\ttfamily\footnotesize,
    breakatwhitespace=false,         
    breaklines=true,                 
    captionpos=b,                    
    keepspaces=true,                 
    numbers=left,                    
    numbersep=5pt,                  
    showspaces=false,                
    showstringspaces=false,
    showtabs=false,                  
    tabsize=2
}

\AtBeginDocument{%
  \providecommand\BibTeX{{%
    \normalfont B\kern-0.5em{\scshape i\kern-0.25em b}\kern-0.8em\TeX}}}

\begin{document}


%\title{MathRAMs: Variable Precision Compute and Memory Blocks for FPGAs}
%\title{MathRAMs: Versatile and Adaptable Compute and Memory Blocks for FPGAs}
%\title{MathRAMs: Fusing Memory and Compute in One Block for FPGAs}

%\title{MathRAMs: Configurable Fused Compute-Memory Blocks for FPGAs}
\title{CoMeFa: Compute-in-Memory Blocks for FPGAs \vspace{-5mm}}

%\title{MathRAMs: Configurable, Fused Compute-Memory Blocks for FPGAs}
%\title{RamCom: Fusing Memory and Compute in One Block for FPGAs}
%\title{MathRAMs for FPGAs: One Hybrid Building Block To Do It All}
%\title{FPGAs Achieve Singularity}


\author{
\IEEEauthorblockN{Aman Arora\IEEEauthorrefmark{1},
Tanmay Anand,
Aatman Borda,
Rishabh Sehgal,
Bagus Hanindhito,
Jaydeep Kulkarni,
Lizy K. John
}
\IEEEauthorblockA{\textit{The University of Texas at Austin} \\
%\textit{}\\
%Austin, United States \\
\IEEEauthorrefmark{1}aman.kbm@utexas.edu
}
%\and
\vspace{-5mm}
}


\begin{figure*}[!t]
\begin{large}
\textcopyright 2021 IEEE. Personal use of this material is permitted.  Permission from IEEE must be obtained for all other uses, in any current or future media, including reprinting/republishing this material for advertising or promotional purposes, creating new collective works, for resale or redistribution to servers or lists, or reuse of any copyrighted component of this work in other works.
\newline
\newline
This work has been accepted at the 2022 30th IEEE International Symposium on Field-Programmable Custom Computing Machines (FCCM) and will appear in the proceedings and on the IEEE website on/around May 15, 2022.
\newline
\newline

\end{large}
\end{figure*}
\pagebreak


\maketitle
\begin{abstract}

Block RAMs (BRAMs) are the storage houses of FPGAs, providing extensive on-chip memory bandwidth to the compute units implemented using Logic Blocks (LBs) and Digital Signal Processing (DSP) slices. 
We propose modifying BRAMs to convert them to \wname~ (\underline{Co}mpute-In-\underline{Me}mory Blocks for \underline{F}PG\underline{A}s) RAMs. These RAMs provide highly-parallel compute-in-memory by combining computation and storage capabilities in one block. 
\wnameRAMs ~utilize the true dual port nature of FPGA BRAMs and contain multiple programmable single-bit bit-serial processing elements. \wname~RAMs can be used to compute in any precision, which is extremely important for evolving applications like Deep Learning. Adding \wnameRAMs ~to FPGAs significantly increases their compute density. We explore and propose two architectures of these RAMs: \wnameD ~(optimized for delay) and \wnameA ~(optimized for area). Compared to existing proposals, \wnameRAMs~do not require changing the underlying SRAM technology like simultaneously activating multiple rows on the same port, and are practical to implement.
\wnameRAMs~are versatile blocks that find applications in numerous diverse parallel applications like Deep Learning, signal processing, databases, etc. By augmenting an Intel Arria-10-like FPGA with \wnameD~(\wnameA) RAMs at the cost of 3.8\% (1.2\%) area, and with algorithmic improvements and efficient mapping, we 
%are able to get upto 6.6x (3.3x) speedup in representative benchmarks.
observe a geomean speedup of 2.5x (1.8x), across several representative benchmarks. 
Replacing all or some BRAMs with \wnameRAMs ~in FPGAs can make them better accelerators of modern compute-intensive workloads.

\end{abstract}
%Compared to similar existing proposals, CoMeFa RAMs are practical, and are not targeted to a specific application like Deep Learning. 

%They have a bit-serial precision-agnostic compute architecture. 
%Using intelligent algorithms and efficient mapping of applications to an FPGA containing \wnameRAMs,
%In our evaluation of various applications, we observe an average speedup of X\% across several representative benchmarks compared to a %LB-heavy
%baseline Intel Arria 10-like FPGA.
%The amount of on-chip storage, in form of BRAMs, has been steadily increasing over FPGA generations. In this paper, 
%Fully programmable architectures like CPUs and GPUs have been deployed as well, but have generally proven . 
%\IEEEpeerreviewmaketitle
%\input{ccs-concepts.tex}
%\input{keywords.tex}

%\fontsize{9.6pt}{11.6pt}\selectfont
\fontsize{9.9pt}{11.8pt}\selectfont

%\vspace{-1mm}
\section{Introduction} \label{introduction}
%\vspace{-2mm}
FPGAs are being used to accelerate workloads ranging from internet search, to baseband processing, to the ubiquitous Deep Learning (DL). FPGAs contain fine-grained programmable logic blocks (LBs), fixed-function math units (DSP slices), and distributed Block Random Access Memory (BRAM) structures that are connected via a highly configurable interconnection fabric. BRAMs play a vital role by storing operands and results on-chip, feeding the compute units with data at a very high bandwidth. %DSP Slices provide hard arithmetic units that can perform computations efficiently, while LBs can be used for arithmetic as well as control.

%We make some observations about 
The current usage paradigms of BRAMs, LBs and DSPs pose limitations to the acceleration that can be achieved using FPGAs. 
In many large FPGAs deployed in cloud applications, hundreds of MBytes of data can be stored on-chip in BRAMs, enabling fully-data-resident acceleration. However, in applications where on-chip storage requirements are low (eg. where data is streamed to the FPGA), BRAMs may be left idle. Additionally, BRAMs on FPGAs support a limited set of heights and widths. 
%For example, BRAMs in Intel FPGAs are 20 Kilobits in size and can be configured as 512x40, 1024x8 or 2048x8 in simple dual port mode (1r,1w), or as 1024x16 or 2048x8 in true dual port mode (2r/w).
%, with 1 or 2 read and write ports. 
This can limit the bandwidth available to the compute units because the data needs to be read out from pins of the BRAM.
%Mismatch in user's requirements and bandwidth obtained 
%In data-parallel applications, where processing thousands of bits of data together is common, designers need wide-and-shallow logical memories. This in turn, means that the data is split over multiple physical BRAMs for higher bandwidth leading to only a few rows of each BRAM being utilized. 
The separation of compute units (LBs and DSPs) from storage units (BRAMs) implies data movement using the routing/interconnect to feed the compute units with input data and to store the outputs back to the storage units. This stresses the routing resources significantly and leads to increased power consumption. 

FPGAs provide the ability to develop hardware for different precisions. This is especially important for DL applications, because the precision requirements change rapidly. DSP slices, however, support a limited set of precisions. 
%Intel FPGAs have supported 18x19 and 27x27 fixed point operations and 32-bit floating point operations, but have recently added support for 9-bit fixed point and 16-bit floating point \cite{intel_agilex}. 
Designers end up implementing low precision math units on LBs instead of DSPs, reducing the number of LBs available for other purposes and leaving DSPs unused.

In this paper, we solve the limitations mentioned above by proposing to convert BRAMs on an FPGA to \wnameRAMs. A \wnameRAM~block enables computation within the RAM array, without transferring the data in or out of it. One-bit bit-serial programmable processing elements are added at the output of the sense amplifiers. This transforms the BRAM into a parallel SIMD (Single Instruction Multiple Data) computation unit. The availability of true dual port mode in FPGA BRAMs~\cite{xilinx_ultrascale_bram}\cite{arria10_overview} is exploited to read operands.
%, instead of changing the underlying operation of the memory array or its peripherals like in similar prior work \cite{ccb}. 

Computation in any precision can be easily performed in \wnameRAMs~without any explicit hardware because it uses bit-serial compute \cite{neural_cache}. \wnameRAMs~reduce the dependence on routing/interconnect and hence increases the routability of the FPGA. Data movement is reduced because the computation is done in the RAM itself, thereby saving power. Since the data is not moved in/out of the RAM block, pin limitations do not restrict the available bandwidth. Instead, the internal physical geometry of the RAM, which is typically higher than port width, governs the effective bandwidth. The compute throughput of the FPGA is increased significantly owing to the massive parallelism that is unlocked because of the existence of numerous RAM blocks on an FPGA. 

\wnameRAMs~are not replacements of DSPs or LBs, but can work together and complement them. 
In some ways, \wnameRAMs~can be thought of blocks that fuse together Logic Blocks and BRAMs. They provide a more structured way of computation compared to Logic Blocks, along with storage capability of BRAMs. They are more flexible  than DSP Slices owing to their precision agnosticism.
They can be used in diverse parallel applications like DL, signal and image processing, databases, compression, encoding, decoding, etc. Because of the bit-serial nature of the compute, \wnameRAMs~are particularly suited for throughput-oriented latency-tolerant workloads. Workloads with low-precision compute and bit-wise operations are also well accelerated using \wnameRAMs. 
Note that \wnameRAMs~can still store data and operate identically like a BRAM. Unused BRAMs can be used as \wnameRAMs~increasing the available compute throughput and hence, achieving faster acceleration. 

%The same operation is performed on in each processing element in a \wnameRAM. 

\begin{comment}

Our contributions in this paper are the following:
\begin{enumerate}
    \item Propose modified BRAM blocks called \wnameRAMs~and describe their architecture and operation. 
    \item Show versatility of \wnameRAMs~for several applications with different workload characteristics.
    \item Quantify the benefit of using \wnameRAMs~for several key metrics and identify the deployment challenges.
\end{enumerate}

\end{comment} 
\section{Background} \label{section:background}

\subsection{Bit Serial Computing}
Bit serial computing is commonly used for digital signal processing and has been used on FPGAs as well \cite{greg_stitt_fpga_bit_serial} \cite{fpga_bit_serial_2}. The main idea is to process one bit of multiple data elements every cycle. This is different from bit-parallel computing, in which multiple bits of one data element are processed every cycle. As an example, a conventional bit-parallel processor will take 128 steps to perform element-wise sum of two arrays with 128 16-bit elements. A bit-serial processor with 128 processing elements would complete the operation in 16 steps as it processes the arrays \textit{bit-by-bit} instead of \textit{element-by-element}. Since the number of elements in arrays is typically greater than the bit-precision of elements, bit-serial computing can provide much higher throughput compared to bit-parallel computing, albeit at higher latency for each result. 

%Still too much information on compute-in-memory. Shrink it to one paragraph

\subsection{Compute-In-Memory}
Compute-In-Memory or Processing-In-Memory (PIM) \cite{pim_workload_perspective} is the paradigm of bringing computation closer to the data, instead of moving data to distant compute units. 
%PIM architectures have been around for decades, but have gained traction in the last decade. This is attributed to innovations such as 3-D stacked memory dies like HBM \cite{hbm} and HMC \cite{hmc}, the ability to perform logic operations using memory cells (in both DRAM and SRAM) and the emergence of resistive memory technologies \cite{magic}. 
Many accelerators using PIM have been proposed and deployed: ReRAM based \cite{isaac} \cite{prime} \cite{floatpim}, DRAM based \cite{drisa} \cite{ambit} \cite{ComputeDRAM}, and SRAM based \cite{compute_memory_uiuc} \cite{x_sram} \cite{dima} \cite{compute_sram}.
%ReRAM-based PIM accelerators like ISAAC \cite{isaac} and PRIME \cite{prime}, FloatPIM \cite{floatpim} have been proposed. %The main challenge of these architectures is integrating them on the same Silicon die that uses a standard CMOS-based process. Only some vertical stacked architectures have been shown to work so far \cite{nature_memristor_cmos}. 
%Samsung recently announced a DRAM product called HBM-PIM \cite{samsung_hbm_pim} which integrates compute units onto a High Bandwidth Memory chip. This paradigm is near-memory compute instead of in-memory compute. Mythic AI's Intelligent Processing Unit (IPU) contains tiles that have analog matrix multiplier which uses Flash memory transistors. Their design requires DACs and ADCs for operation.
Computational RAM (or C-RAM) \cite{computational_ram} is an architecture where a row of processing elements (PEs) is added to a memory (DRAM or SRAM) to convert it into a SIMD processor.
%, as shown in Figure \ref{fig:c_ram_architecture}. 
%Each PE is pitch-matched with a memory column. 
%An instruction is received by the memory from the host, operand rows are read and stored in the PEs, the operation is then performed, and the results are stored back into a row. All PEs in a memory execute the same instruction in a cycle. 
%Two PE architectures were proposed - A simpler PE supporting bit-serial computation and a more complex PE with bit-parallel support. 
%This is shown to achieve significant speedup for applications like image processing, databases, computer-aided design, etc. 
Jeloka et. al. \cite{supreet_logic_in_memory} created a logic-in-memory SRAM prototype 
%shown in Figure \ref{fig:logic_in_memor_technology}. 
where multiple word lines are activated simultaneously and the shared bit-lines can be sensed, effectively performing logical AND and NOR operations on the data stored in the activated rows. 
%To avoid data corruption because of multi-row access, the wordline voltage has to be lowered significantly. This is required to ensure robustness and high margins. An additional row decoder is required for multi-row access. Additionally, in this architecture, one sense amplifier for each pair of bitlines (BL/BLB) is replaced with two sense amplifiers (one with BL/Vref and second with BLB/Vref). 
%This technology is deployed on CPU caches to transform them into parallel processing engines \cite{compute_cache}.
%, leading to speedups in many applications involving operations like word count, string match, etc. 
In Neural Cache, Eckert et. al. apply this technology to DL applications \cite{neural_cache}, adding processing elements to the sense amplifiers and deploy bit-serial compute to perform DL operations. 
%Elements are stored in a transposed data layout - all the bits of one data word are stored vertically in cells on the same bitline, such that bit i of all the data words map to the same wordline. 
Wang et. al. proposed integrating the technology from Neural Cache into FPGA BRAMs to create Compute Capable BRAMs (CCB) \cite{ccb}. 
%They show speedups for recurrent neural networks (vanilla RNN, LSTM and GRU), for int8 and 8-bit block floating point precisions. 
%This architecture needs one sense amplifier and write driver per column (no column multiplexing).
The complexity associated with the reduction in voltage required to ensure robustness when activating multiple wordlines makes this architecture not very practical.

\begin{comment}

\begin{figure}[hbt!]
\centering
\includegraphics[width=0.9\linewidth]{figures_bagus/c-ram-architecture.pdf}
\caption{Architecture of C-RAM \cite{computational_ram}}
\label{fig:c_ram_architecture}
\end{figure}

\end{comment}

\begin{comment}

\begin{figure}[hbt!]
\centering
\includegraphics[width=0.9\linewidth]{figures_bagus/logic_in_memory_technology.pdf}
\caption{Logic-in-memory technology from Jeloka et. al. (for a single port RAM) \cite{supreet_logic_in_memory}}
\label{fig:logic_in_memor_technology}
\end{figure}

\end{comment}

%\subsection{FPGA storage and compute}
%Compute units on FPGAs are designed using soft logic (LBs) or hard logic (using DSP Slices). FPGA BRAMs are the on-chip storage houses of BRAMs. They do not have any compute capability in them. BRAMs have programmable height and widths and dual ports because they are commonly used to build FIFOs. BRAMs in Intel FPGAs are 16 Kilobits in size (ignoring ECC bits) and can be configured as 512x32, 1024x16 or 2048x8 in simple dual port mode (1r,1w), or as 1024x16 or 2048x8 in true dual port mode (2r/w). Intel and University of Michigan researchers recently proposed integrating the technology from Neural Cache into FPGA BRAMs to create Compute Capable BRAMs (CCB) \cite{ccb}. They show speedups for recurrent neural networks (vanilla RNN, LSTM and GRU), for int8 and 8-bit block floating point precisions. 

\section{Proposed Architecture: \wnameRAMs} \label{section:architecture}

In this section, we describe the proposed changes to BRAMs to convert them to \wnameRAMs. We consider a BRAM size of 20 Kilobits as in the modern Intel FPGAs, with support for single port, simple dual port and true dual port modes, with the 512x40 being the shallowest and widest configuration. This BRAM has a physical geometry of 128 rows x 160 columns with a column multiplexing factor of 4 \cite{arch_enhancements_in_stratix_v} \cite{arria_10_arch}. 
%The BRAM supports both simple dual port mode with 512x40, 1024x20, and 2048x10 configurations supported and true dual port modes with 1024x20 and 2048x10 configurations supported.

\begin{figure}[t]
\centering
\includegraphics[width=0.8\linewidth]{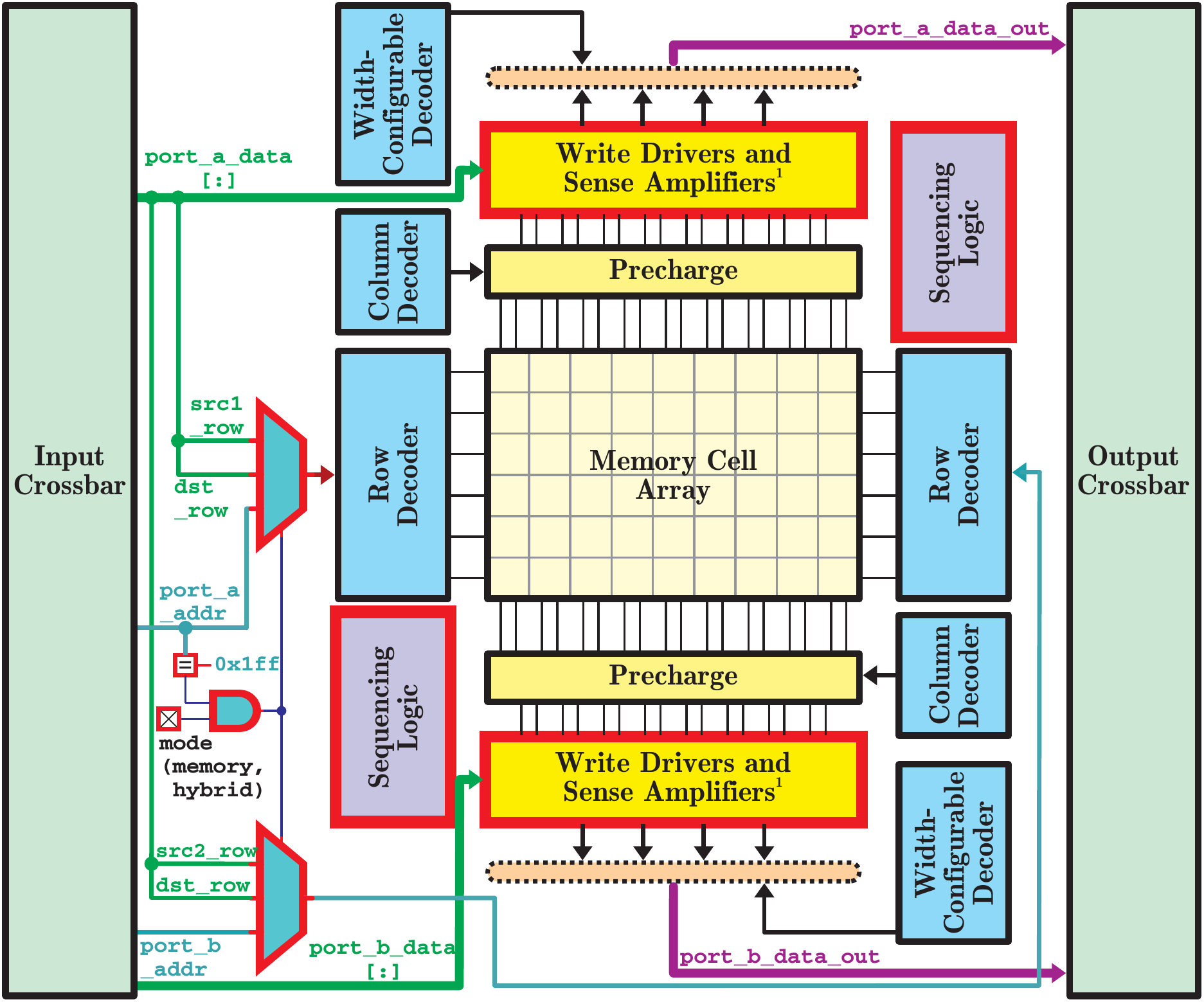}
%\vspace{-2mm}
%\scriptsize
%    \begin{flushleft}
%	\item $^1 CHANGE ME!!!!$ 
%	\end{flushleft}
%\vspace{-3mm}
\caption{Top-level logical diagram of an FPGA BRAM \cite{dont_forget_the_memory} with added/modified blocks for \wnameRAM~highlighted in red. Sense amps of the two ports are far apart, but in physical layout, they are adjacent to each other. This ensures practicality of adding a set of PEs fed by both set of sense amps.}

\label{fig:top_level_diagram}
\vspace{-2mm}
\end{figure}

\subsection{Overview}
%FPGA BRAMs are true dual ported RAMs. They have an 8-transistor SRAM cell, instead of the traditional 6-transistor SRAM cell found in single port RAMs. So, we can read two rows at the same time by providing appropriate row addresses on the ports.

%Utilizing the true dual ported nature of the BRAM, we read two rows at the same time. The data read from these two rows act as operands for the PE that is connected to the output of sense amplifier. Computation happens in the PE and the result is written back to a third row. 

Figure \ref{fig:top_level_diagram} shows a top-level diagram of an FPGA BRAM, with blocks modified/added for \wnameRAM~shown with a red outline. We explore two architectures at the ends of the area-delay design space (evaluating other candidates in this space is left as future work). 

%Note that the operation of the main array is not modified at all, as mentioned in Section \ref{section:related_work}. 

\textbf{\wnameD}: In this architecture, we add additional sense amplifiers and write drivers to enable reading and writing a row in all columns (bitline pairs) together. A processing element (PE) is added below each column. This is similar to the architecture used in \cite{neural_cache}, \cite{compute_cache} and \cite{ccb}. During physical design/implementation, PEs should be laid out so that they pitch-match with the SRAM cells (and sense amplifiers and write drivers) for a bitline pair (BL and BLB). There are 160 sense amplifiers and write drivers per port, and 160 PEs. This provides a parallelism 160 operations done in 1 clock cycle (slightly longer than the baseline BRAM's clock period) at cost of high area overhead. \textbf{This architecture is more practical than CCB because multiple wordlines are not accessed simultaneously on a port and voltage reduction is not required for robustness}.

\textbf{\wnameA}: In this architecture, the number of sense amplifiers and write drivers stays the same as the baseline. A different PE is added below each multiplexed column. An optimization technique called sense amp cycling \cite{cache_automaton} is employed to sequentially sense column multiplexed bits in an extended clock cycle. There are 40 sense amplifiers and write drivers per port, and 40 PEs in the RAM. This provides a parallelism 160 operations done in 1 extended clock cycle, thereby trading off delay for area. \textbf{This architecture has the highest practicality among CCB and \wname~variations.}

%Note that these two architectures are the extreme ends of the area-delay design space and architectures in between (for example, with 80 sense amps i.e. column multiplexing of 2) could be explored. 

The PEs use wires from the peripheral circuitry of both ports of the RAM.
%Although in logical diagram of Figure \ref{fig:top_level_diagram}, the peripheral circuitry of the two ports of the RAM (decoders, write drivers, sense amplifiers, etc) are shown in diagrammatically opposite parts of the figure, in a typical physical layout of a RAM block, they are close to each other, so it is practical to make this modification. Note that PEs are shared between both ports.
%This implies that our proposed modifications are realistic.
Sequencing logic that sequences the events of the read/write operations (wordline activation, precharge, sense amp enable, etc) in the memory is modified. This is done to support doing both read and write in one cycle. 
Some additional logic (comparator, configuration bit, multiplexers in front of row decoders) are also added, and will be explained in detail in this section.

\subsection{Modes and Phases}

As shown in Figure \ref{fig:top_level_diagram}, a new configuration SRAM cell is added which decides the mode of operation of a \wnameRAM~block. 
%This cell is not modified dynamically during operation (we leave considerations related to dynamic partial reconfiguration as future work). 
A \wnameRAM~can operate in two modes:
\begin{itemize}
    \item \textbf{Memory Mode}: In this mode, \wnameRAM~behaves as a conventional BRAM with no change in functionality. In this mode, the designer can flexibly configure the number of ports and the width/depth of the BRAM.
    \item \textbf{Hybrid Mode}: If this mode is enabled at configuration time, the \wnameRAM~can be used for computation as well as storage. In this mode, the RAM is automatically configured to its maximum width (512x40) to maximize the read/write throughput for populating the memory array with input data and reading the results. A special address (0x1FF) is reserved. A comparator is added to Port A's address signal to check for this address (Figure \ref{fig:top_level_diagram}). Data written to this address is treated as an instruction. Accessing other addresses is done normally; used for storing operands and reading results. 
\end{itemize}

%To perform any compute operations using a \wnameRAM, it should be configured in Hybrid mode at FPGA configuration time. 
\begin{comment}

Operations on \wnameRAMs~typically happen in 3 stages: 
\begin{itemize}
    \item \textbf{Data loading stage}. Input data is stored in transposed format into the memory array in this stage. Any address other than 0x1FF can be written to, in this stage.
    \item \textbf{Compute stage}. In this stage, instructions are provided to \wnameRAM~by writing to address 0x1FF. Source operand rows are read, processing elements evaluate the results and results are written to destination rows. 
    \item \textbf{Data unloading stage}. Results can be read out in this stage by reading them from any address in the memory array.
\end{itemize}
\end{comment}

A clock cycle during computation has 3 phases. In the first phase, two rows containing operand bits are read by activating the corresponding word lines, one from each port. In the second phase, the logic gates in the PE compute the result. In the third phase, the result is stored back by activating a wordline. This leads to a longer clock period, compared to typical BRAM. 
%However, the final clock period is shorter in \wnameRAM, compared to CCB, owing to no reduction in wordline voltage.

\subsection{Processing element} \label{processing_element_arch}
\begin{figure}[t]
\centering
\includegraphics[width=0.8\linewidth]{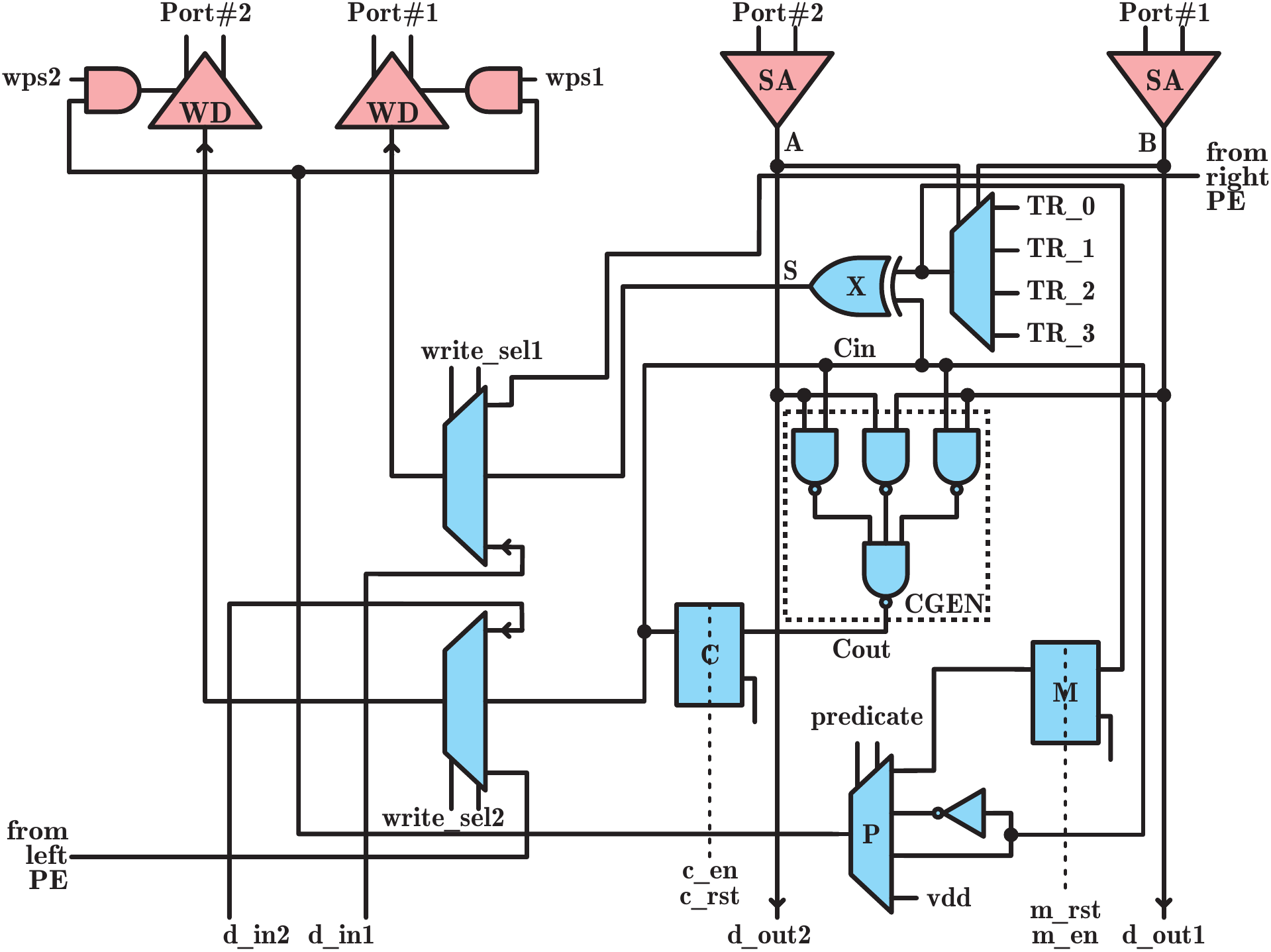}
\caption{Architecture of the processing element used in \wname-D}
\label{fig:processing_element_comefa1}
\vspace{-2mm}
\end{figure}

Fig \ref{fig:processing_element_comefa1} shows the structure of PE added to each column of the memory in \wnameD. On the read path, \textbf{A} and \textbf{B} are the bits of the two operands read from the memory at sense amplifiers \textbf{SA1} and \textbf{SA2} of the two ports. Multiplexer \textbf{TR} evaluates a logical function of \textbf{A} and \textbf{B}, depending on the inputs \textbf{TR0, TR1, TR2, TR3} (truth table) that come from the instruction. If a 2-bit addition is required, the truth table bits will correspond to that of an XOR gate. The output of \textbf{TR} goes through another XOR gate (\textbf{X}) to generate the addition of the input bits including the previous cycle's carry (\textbf{S}). Gates to generate the carry (\textbf{CGEN}) are also present. The \textbf{carry} is stored in the carry latch (\textbf{C}) and can be used in the following cycle's computation. If an addition operation is not required, the carry latch is reset with \textbf{C\_RST}=1, which enables \textbf{X} to pass the output of \textbf{TR} transparently to the \textbf{S} wire. \textbf{C\_EN}=0 disables the latch so it keeps the old value. The read outputs \textbf{A} and \textbf{B} are also sent to \textbf{d\_out1} and \textbf{d\_out2}, which is the normal read path.

\begin{figure}[t]
\centering
\includegraphics[width=0.9\linewidth]{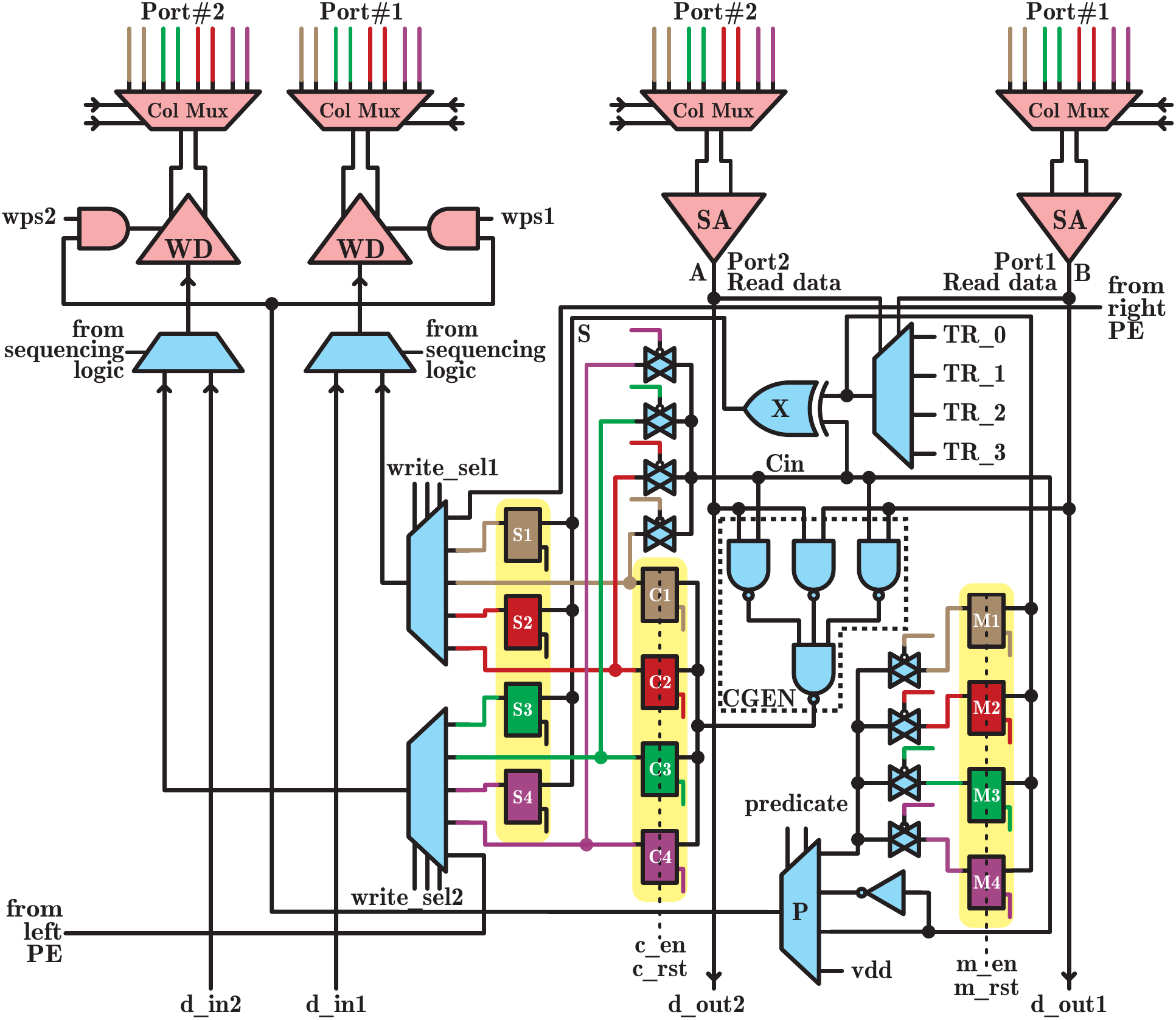}
\caption{Architecture of the processing element used in \wname-A}
\label{fig:processing_element_comefa2}
\vspace{-2mm}
\end{figure}

On the write path, 3-input multiplexers \textbf{W1} and \textbf{W2} are added before the write drivers of the two ports. These multiplexers determine the sources for the write operation. \textbf{W1} can select between the \textbf{S}, the input data port \textbf{d\_in1} (normal write operation) and the value read from the right neighboring PE (used during left shift operation). \textbf{W2} can select between the \textbf{carry}, the input data port \textbf{d\_in2} (normal write operation) and the value read from the left neighboring PE (used during the right shift operation). The mux select signals for \textbf{W1} and \textbf{W2} come from the instruction. 

The output of multiplexer \textbf{TR} is also stored in a special latch called \textbf{M} and is called \textbf{mask}. Predication logic allows enabling/disabling the write drivers (\textbf{WD1} and \textbf{WD2}). For this, a multiplexer (\textbf{P}) is added to select the signal that will enable/disable the write drivers. The \textbf{mask}, \textbf{carry}, \textbf{not-carry} and \textbf{VDD (logic 1)} can be selected. This helps \wnameRAMs~mask writing the results based on various conditions, like the value of the \textbf{mask} or the \textbf{carry} bit, to support multiplications and floating point operations. The \textbf{wps*} signals decide which port's write path is activated for a given cycle. 
%The \textbf{port} signal decides which port's write path is activated for a given cycle.

\begin{figure}[t]
\centering
\includegraphics[width=0.8\linewidth]{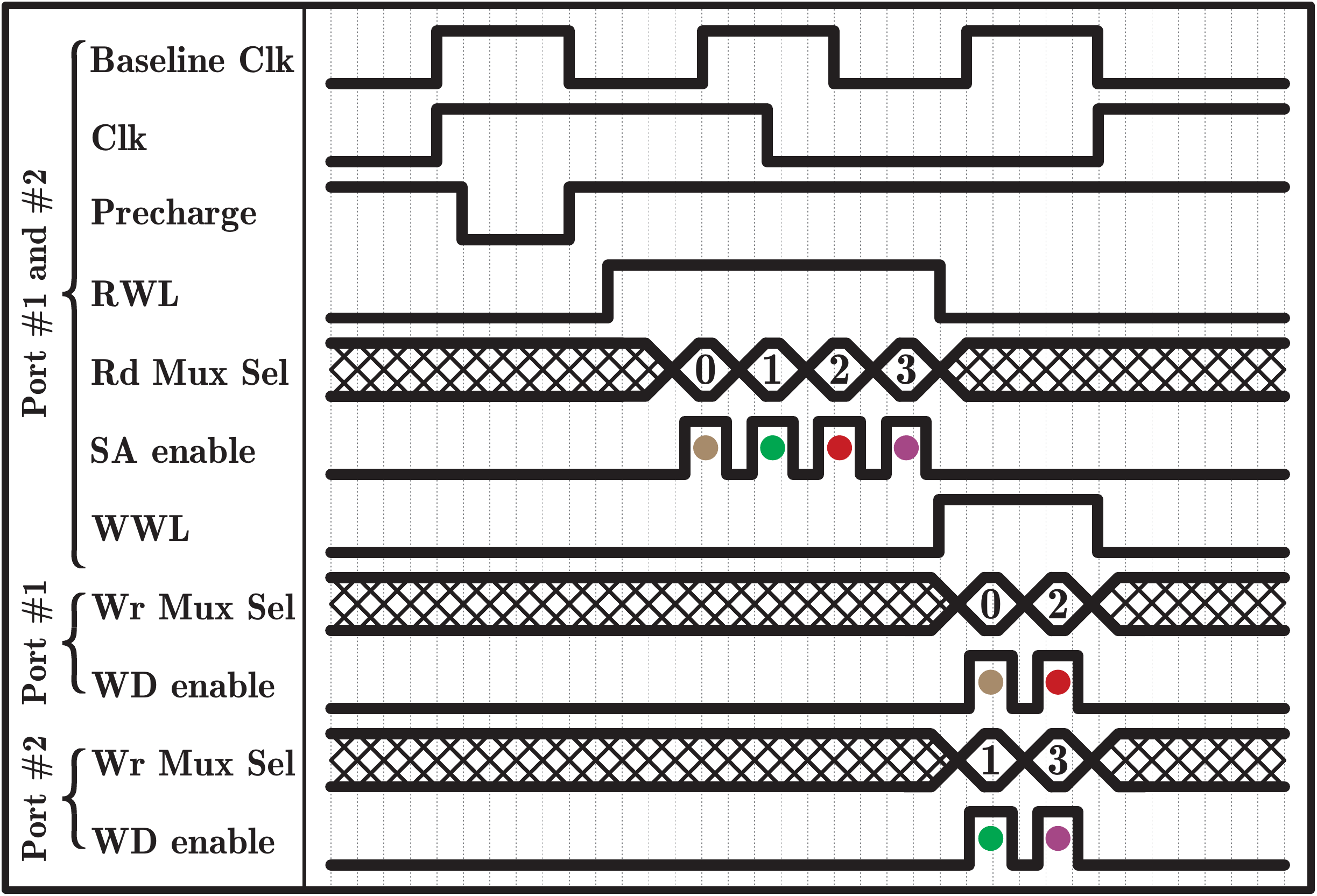}
\caption{Sequence of operations in one clock cycle of \wname-A}
\label{fig:read_write_comefa_2}
\vspace{-2mm}
\end{figure}

Fig \ref{fig:processing_element_comefa2} shows the structure of PE added to each multiplexed column of the memory in \wnameA.  All the labels have the same meaning as the PE described above. The number of \textbf{C} and \textbf{M} latches changes to 4, and there are 4 additional latches for S. On each port, 4 column-multiplexed bits are read and two results are written back in an extended clock cycle. In the read phase of the cycle, the brown bitline pairs from each port are sensed first. The resulting \textbf{S} bit is stored in latch \textbf{S1}, carry bit \textbf{C} is stored in the latch \textbf{C1}, and mask bit \textbf{M} is stored in latch \textbf{M1}. This is repeated for red, green, and purple bitline pairs successively. All \textbf{Sn}, \textbf{Cn} and \textbf{Mn} latches get updated in this process. Then, in the write phase of the cycle, results for the brown and red bitlines are written using the write drivers of the two ports, followed by the green and purple ones. This is shown in Fig \ref{fig:read_write_comefa_2}. Clocks in the PE are driven by signals derived from the sense amplifier enable pulses. The paths in the PE do not add any additional delay to the extended clock from sense amp cycling.

\vspace{-1mm}
%\vspace{-1mm}

\subsection{Instructions}
\begin{figure}[h]
\centering
\includegraphics[width=\linewidth]{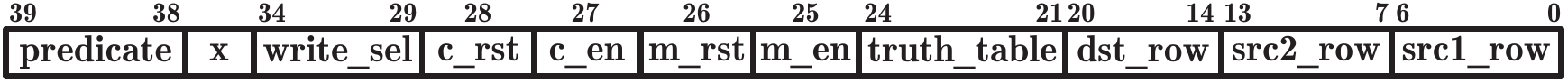}
\caption{Instruction format for \wnameRAMs}
\label{fig:instruction_format}

\end{figure}

An instruction is 40-bits and is only required to be written to Port A's data bus. The format is shown in Figure \ref{fig:instruction_format}.
The field names in the instruction are self-explanatory. They directly drive the corresponding signals in the PE (e.g. \texttt{predicate} bits are applied to the select lines of the multiplexer \textbf{P}).
%in the PE. The \texttt{write\_sel} bits are used to derive the select lines of the multiplexers \textbf{W1} and \textbf{W2}. 
%\texttt{C\_EN} and \texttt{M\_EN} are applied to the enable signals of the carry latch (\textbf{C}) and mask latch (\textbf{M}). \textbf{C\_RST} and \textbf{M\_RST} are applied to the reset signals to these latches. The \texttt{truth\_table} bits are inputs to multiplexer \textbf{TR} in the PE. 
The \texttt{src1\_row}, \texttt{src2\_row} and \texttt{dst\_row} bits are used for activating first operand row on Port A, second operand row on Port B and the row at which results will be stored, respectively. These addresses are fed to the appropriate row decoders at the right time in the clock cycle by the sequencing logic in the \wnameRAM~using the multiplexers shown in Figure \ref{fig:top_level_diagram}.
A designer can choose to generate instructions by finite state machines (FSM) implemented in soft logic, or store them in another BRAM on-chip and apply them to a \wnameRAM. Multiple \wnameRAMs~can share instruction generation logic to amortize its cost. 
%An alternative implementation is to harden the instruction storage and decoding inside a BRAM as well. This will increase the size of the BRAM significantly, and also reduce the flexibility. We leave detailed exploration of this as future work.

%\begin{figure}[t]
%\centering
%\includegraphics[width=0.7\linewidth]{figures_bagus/matram_operation.pdf}
%\caption{Operation of \wnameRAM~shown for 4-bit %operands and a 4-bit result}
%\label{fig:math_ram_operation}
%\vspace{-2mm}
%\end{figure}

\vspace{-4mm}
\subsection{Operation} \label{section:operation}

All computation is done in a bit-wise manner, using transposed data layout. Figure \ref{fig:math_ram_ops_shift_architecture} (a) shows how elements are stored, read, computed on (based on the \texttt{truth\_table} bits in the instruction) and the result stored back, row-by-row.
%Consider an example of bit-wise \texttt{AND}ing of the elements of two arrays (array length = 160 and element width = 4 bits).  Each element is stored in a column, 1 bit in 1 row. Elements of array 1 are stored in rows $i$, $i+1$, $i+2$ and $i+3$. Elements of array 2 are stored in rows $j$, $j+1$, $j+2$ and $j+3$. A total of 8 rows and 160 columns are required to store both arrays. In one cycle, rows $i$ and $j$ are read, each PE computes the \texttt{AND} of two bits (based on the \texttt{truth\_table} bits in the instruction), and the result is stored in row $k$. This process is repeated 4 times with increasing row addresses, and the final result is available, after 4 cycles, in rows $k$, $k+1$, $k+2$ and $k+3$. This is shown in Figure \ref{fig:math_ram_operation}. 
%Note that 128 operations are done in parallel in each cycle in the \wnameRAM.
%This example case can be removed or just a table (like the one used in MathRAMs can be rather presented (to show cycles for add and mult operation cycles. 
%Other operations follow a similar data layout and operation mechanism. 
For performing addition operation, in each cycle, the PE computes the sum, stores the carry for the next cycle computation, and writes back the sum bit to the result row. Then the next bit position is processed in the next cycle, using the stored carry. The final carry is stored into a row using an extra cycle. Thus, the addition for $n$-bit operands takes $n+1$ cycles. 
Multiplication is based on iterative addition of partial results. In each iteration, one bit of the first operand is loaded as into the \texttt{mask} latch, and the second operand’s bits are added to the partial sum only if the \texttt{mask} is 1. Multiplication of $n$-bit operands takes $n^2 + 3n -2$ cycles.
The addition and multiplication algorithm mentioned above, and also in-RAM reduction algorithm are detailed in \cite{neural_cache}.
%Reduction operation across operands stored in columns is done by shifting the operand bits. 

%\subsection{Support for Shift Operations}
\subsection{RAM-to-RAM chaining} \label{section:chaining}
\begin{figure}[t]
\centering
\includegraphics[width=\linewidth]{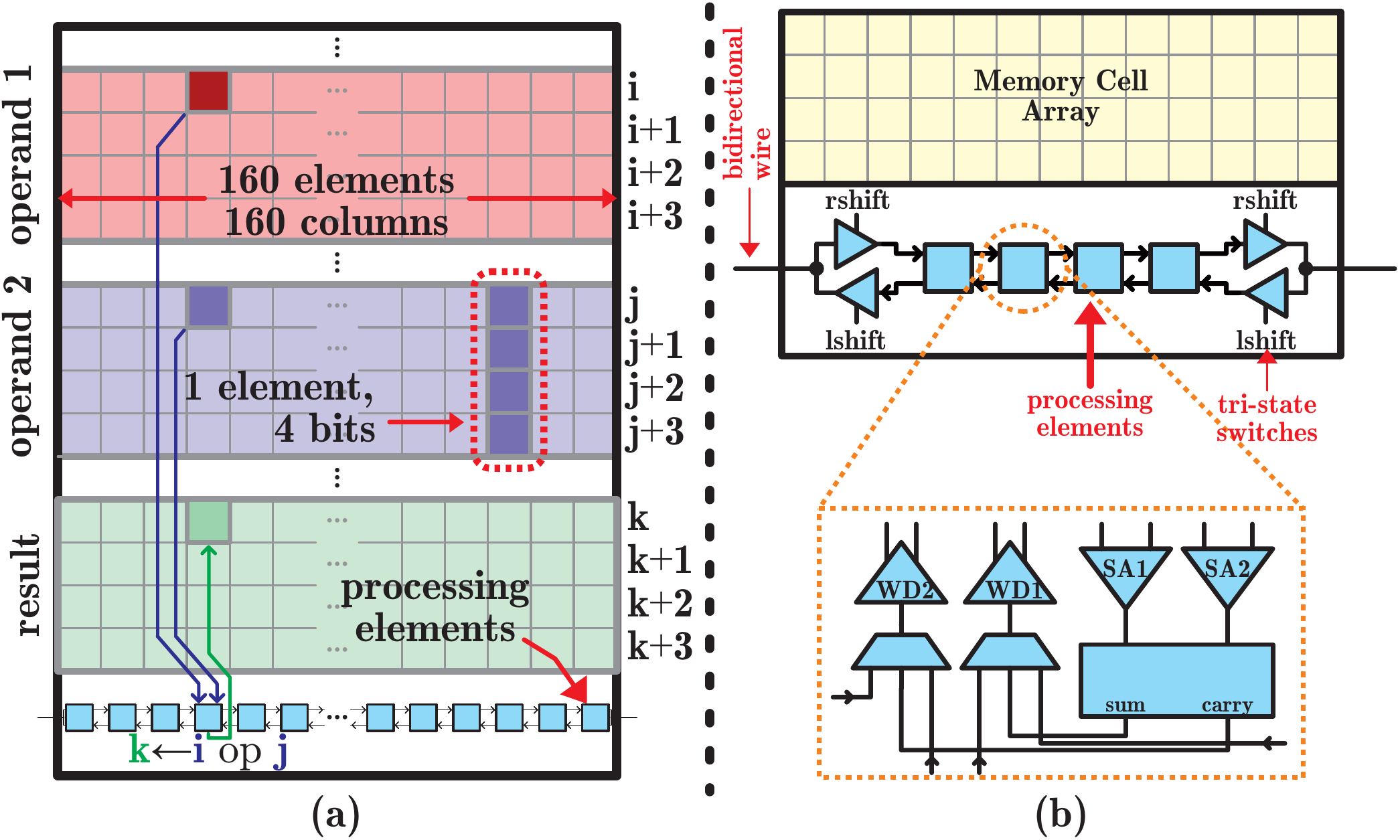}
\caption{(a) Operation of \wnameRAM~shown for 4-bit operands and a 4-bit result. (b) \wnameRAM~supports shifting within a block and across blocks using chaining}
\label{fig:math_ram_ops_shift_architecture}
\vspace{-2mm}
\end{figure}

\wnameRAMs~provide the capability of performing left-shift and right-shift operations efficiently.
%, through the \texttt{write\_sel} bits in the instruction and the values from neighboring PEs applied to the \textbf{W1} and \textbf{W2} multiplexers. 
Shifts are single operand operations. For a left (right) shift operation, the source operand row is read into the PEs, each PE's \textbf{W1} (\textbf{W2}) mux is configured to select the bit read from the right (left) neighboring PE, and that bit is written into the destination row. For \wnameA, shifting values from a bitline pair to another bitline pair within the same column multiplexer is also supported.
%done by decoding the \texttt{write\_sel} bits of the instructions and setting the select lines of \textbf{W1} and \textbf{W2} muxes appropriately.
Direct connections are provided between neighboring \wnameRAMs~in a column to allow for shifting data between the corner PEs in each \wnameRAM.
%~(PE 127 of one \wnameRAM~and PE 0 of the next \wnameRAM~and so on). %These are similar to carry chains in Logic Blocks and cascade chains in DSP Slices in modern FPGAs. Xilinx FPGAs have direct BRAM-to-BRAM interconnections as well.
These connections can provide a much easier way to perform inter-\wnameRAM~communication and obtain even more parallelism. Figure \ref{fig:math_ram_ops_shift_architecture} (b) shows these direct connections between \wnameRAMs~along with the details of the shift operation support inside each PE.
%If unidirectional wiring is used, four additional pins would be required on the \wnameRAM~hard block to allow for shifting in both directions (input from top block, output to bottom block, input from bottom block and output to top block). To minimize this overhead, we choose to provide bidirectional wires controlled by tri-state switches, because at one time, shifting in only one direction is required. The tri-state switches are controlled by a signal decoded from the \texttt{write\_sel*} and \texttt{wps*} bits, because they govern the shift direction.

\vspace{-2mm}
%\subsection{Support for floating point operations}
%\subsection{Precision Agnosticism}
\subsection{Variable precision support}
%PEs in \wnameRAM~consist of low-level digital logic elements and there is 
Hardware in \wnameRAM~PEs is not specific to any numerical precision. 
%The bit-serial nature of operations can be used to perform operations in any numerical precision.
A different sequence of instructions is all that's required to compute in a different precision. 
%This is especially important for Deep Learning applications, where custom smaller numerical precisions are commonly used. 
The sequences for fixed-point addition and multiplication were explained in Section \ref{section:operation}. \wnameRAMs~can natively support floating point precisions as well, as opposed to CCB \cite{ccb}. We adapt the floating point algorithms for addition and multiplication from FloatPIM \cite{floatpim}. The \wnameRAM~PE can perform all the steps in the sequences because: (1) \textbf{carry} and \textbf{not-carry} are used in the predication logic, (2) \textbf{mask} is populated from the output of the programmable multiplexer \textbf{TR} instead of just \textbf{A} or \textbf{B}, and (3) operations like XOR can be performed easily using \textbf{TR} and the \texttt{truth\_table} fields in the instruction. 
%A challenge in performing floating point operations is that the values in different columns of a \wnameRAM~can have different exponents and so the same steps can not be applied to all columns for full parallelism. The predication logic in the PE enables masking the operations for some columns that do not satisfy a condition. 
%Another interesting feature we use for floating point operations is to have one or more rows containing constants (1/0) and use them as operands, for example, for handling the implicit 1 in the mantissas. This can be done during data loading or on-the-fly by programming the truth table for an instruction such that it results in a constant. 
The approximate number of cycles consumed for floating point multiplication and addition are $M^2+7M+3E+5$ and $2ME+9M+7E+12$, where $M$ = number of mantissa bits and $E$ = number of exponent bits.

\vspace{-1mm}
\subsection{Data loading and unloading}
\begin{figure}[t]
\centering
\includegraphics[width=0.8\linewidth]{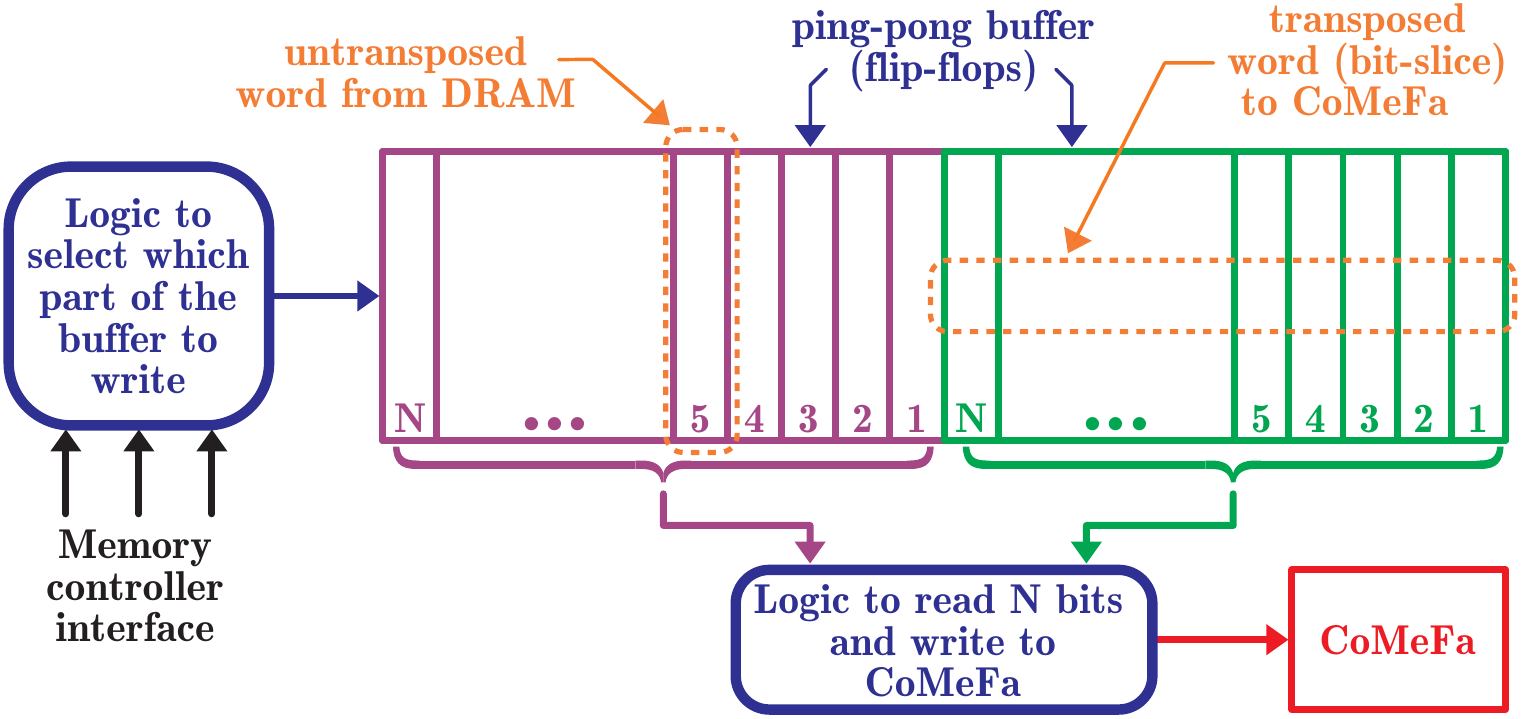}
\caption{Swizzle logic to load non-transposed data from DRAM directly into \wnameRAM~in transposed layout (N=40)}
\label{fig:swizzle_logic}
\vspace{-2mm}
\end{figure}

As mentioned in Section \ref{section:operation}, data has to be stored in a transposed layout in \wnameRAMs~for computation. We design a \textbf{swizzle module} (implemented in soft logic) that can be used to read data from DRAM, transpose it and write it a \wnameRAM~on-the-fly. The architecture of the swizzle logic is shown in Figure \ref{fig:swizzle_logic}. The swizzle module employs a ping-pong buffer FIFO. Untransposed data read from DRAM is written in-order into the ping part of the FIFO (depth = 40 elements). When the ping part is full, a transposed word (a bit slice from 40 elements) can be read and written into consecutive bitlines on the same wordline in a \wnameRAM, and new data from DRAM is written into the pong part. 
%After the pong part is full, transposed data is read from the pong part and written into a \wnameRAM, and new data from DRAM is now written into the ping part again. 
%This process continues until the required data has been populated into \wnameRAMs. 
In a similar fashion, transposed data can be read from \wnameRAMs~and stored into DRAM in untransposed form using swizzle logic. 
%Swizzle logic is implemented in soft logic and the number of instances required are typically governed by the DRAM bandwidth available on chip.

\vspace{-1mm}
\subsection{One Operand Outside RAM (OOOR) operations}\label{section:ooor}
In Section \ref{section:operation}, two operands were stored inside the RAM. However, in many cases, an optimization can be applied - one of the operand can be outside the RAM. E.g, multiplying an array of numbers (stored in the RAM) with a scalar operand (outside the RAM). We call these OOOR ops. 
This method saves space inside the RAM. Without OOOR, in the multiplication example, we would need to replicate the scalar operand in each column.
It allows easy inspection of outside operand's bits, thereby enabling efficient algorithms. For example, in the normal shift-and-add based multiplication explained in \cite{neural_cache}, if a bit in the scalar operand is 0, cycles are still consumed, which can be avoided by using OOOR. In the average case, half of the bits will be 0 and therefore, the number of cycles can be reduced by 50\%. Efficient algorithms like booth multiplication can also be deployed. We apply OOOR to dot-product where one of the vector’s elements are common to all columns. Looking at corresponding pairs of bits across the operands outside the RAM and adding partial sums inside the RAM based on their values enabled a 2x speedup compared to the naive algorithm. 
 %This is useful in many applications like matrix-vector multiplication and 1-D convolution (FIR filter).
Overall, OOOR operations make the processing elements more powerful and capable by enabling expressing 2 (or 3) operand operations as 1 (or 2) operand operations for the PE.

\vspace{-1mm}
\section{Evaluation} \label{section:evaluation}

\subsection{Tools and Methods Used}
The Verilog-to-Routing (VTR) tool flow \cite{vtr8} is used to evaluate and compare FPGA architectures. 
%VTR synthesizes and implements a given benchmark design on the given FPGA architecture and generates resource usage, area and timing reports. 
%VTR takes two inputs. The first input is an architecture  description  file containing information about an FPGA's building blocks and interconnect resources.  The second input is a Verilog design.  VTR synthesizes and implements the design on the given FPGA architecture and generates resource usage, area and timing reports.
%CCB \cite{ccb} uses Intel Quartus for their experiments, which only provides approximate results because the CCB block is approximated using a BRAM block. 
%with the provided architecture, 
To obtain the area and delay values for the various components of the FPGA, including \wnameRAMs, 
%(to enter them in the FPGA architecture description file for VTR), 
we use COFFE \cite{coffe2}.
%, a transistor sizing and modeling tool for FPGAs. 
%COFFE supports generating and optimizing circuitry for all FPGA components including logic blocks, routing, and hard blocks like BRAMs and DSP slices. It uses Synopsys HSPICE for SPICE simulations, Synopsys Design Compiler for synthesis, Cadence Encounter for placement \& routing, and Synopsys Primetime for timing analysis.
%Based on the provided FPGA architecture properties, it performs SPICE simulations to iteratively optimize the transistor sizes and generates areas and delays in various subcircuits in the FPGA. COFFE seamlessly supports generating and optimizing BRAM circuitry as well. COFFE also supports a hybrid flow in which the core of blocks like DSP Slices is implemented using a standard cell flow and the interface to the interconnect (local crossbar, switch box, connection box, etc) is implemented in full custom using SPICE. The standard cell flow uses Synopsys Design Compiler for synthesis, Cadence Encounter for placement \& routing, and Synopsys Primetime for timing analysis.
COFFE based SPICE simulations use 22nm libraries from Predictive Technology Model \cite{asu_ptm}.
We also perform SPICE simulations using FreePDK45 \cite{ncsu_freepdk45} to get more confidence that the read+compute+write operation of \wnameRAMs~works and to validate the numbers obtained from COFFE. 
%Our standard cell library was the 45nm GPDK library from Cadence. We used scaling factors from Stillmaker et al. \cite{Stillmaker201774} to scale down from 45nm to 22nm. 
%Synopsys VCS and Xilinx Vivado's integrated simulator are used for functional verification. 
%When running COFFE, we used a cost factor of $area * delay ^2$ as it reflects the greater emphasis on delay compared to area, which is typical of high-performance FPGAs like Stratix 10.

We use an analytical model to estimate the energy consumption. We add transistor energy and wire energy. For transistor energy, we use an activity factor of 0.1 and calculate the energy based on the number of transistors in each block (obtained from the area consumed by the block from VTR). For wire energy, we use wire energy numbers (fJ/mm) from \cite{gpus_parallel_computing}, scale them to 22nm technology node using \cite{Stillmaker201774} and multiply that with the total routing wirelength from VTR.

\subsection{Baseline vs. Proposed Architectures} \label{section:baseline_vs_proposed_arch}

%Columnar architecture
%Islnad style layout

%The goal of evaluating \wnameRAMs~is to compare the performance of an FPGA with \wnameRAMs, LBs and DSP slices, with an FPGA with only the traditional building blocks (BRAMs, LBs, and DSP slices). 
We use an Intel Arria 10-like FPGA architecture as baseline with the same resources as Aria 10 GX900 \cite{arria10_product_table} (Table \ref{table:baseline_fpga_resources}. Arria 10 FPGAs \cite{arria10_overview} use a  technology node (20nm) similar to our setup (22nm). Arria 10 GX900 has 96 transceiver channels that support upto 17.4 Gbps \cite{arria10_phy_ug}. We assume that a 4-port full-width soft HMC (Hybrid Memory Cube) controller \cite{hmc_user_guide_arria10} is implemented on the FPGA to provide a DRAM bandwidth of 2048 bits/clock. Resources consumed by the controller are not used to map the applications to the FPGA.
%The applications mapped to the FPGA use the total resources available minus those used by the controller. 

%There are differences between our baseline architecture and Intel Stratix 10 (e.g. we do not model HyperFlex). Also, in our evaluation, we use 22nm technology node, but Intel Stratix 10 devices are 14nm. But the results will hold for any FPGA architecture as long as the baseline and the proposed FPGA architectures only differ in presence/absence of \wnameRAMs. 
%To ensure a faster turnaround time while running experiments, we use a shrunk version of the largest Stratix 10 GX 2800 FPGA. We proportionately reduce the available resources (LBs, BRAMs, DSPs) and also the DRAM bandwidth. We assume 70\% efficiency in DRAM bandwidth for our analysis, to ensure that the results are realistic.

%This FPGA has the same resources Stratix 10 GX 650 FPGA. 

We use the VTR FPGA architecture used in \cite{koios}, modified for our requirements, to make a baseline architecture file.
%We start with the VTR FPGA architecture used in \cite{koios} which is a modern island-style architecture where the building blocks are arranged in columns. 
%This architecture has areas and delays of various blocks from COFFE 22nm simulations.
%and has switchboxes from Stratix IV FPGA. 
%We define the grid to approximate the number of blocks (LBs, DSPs, BRAMs) used by GX650.
We run COFFE simulations on an Arria-10 like DSP to indentify its delay and area.
We get delay and areas of a 20 Kilobit BRAM from COFFE (by interpolating between 16K and 32K). 
We scale these results based on the DSP and BRAM delays specified in \cite{arria10_datasheet}. The DSP slice works at 630 MHz in fixed-point mode and 550 MHz in floating-point mode. The BRAM works at 735 MHz in single port, simple dual port and true dual port modes.
%, similar to Arria 10. 
%This gives us the baseline architecture file.
The proposed architecture files (\wnameD~and \wnameA) differ from the baseline in having \wnameRAMs~instead of normal BRAMs. 
%For this, we define an additional mode (Hybrid Mode) in the BRAM model of the baseline architecture. 
%We evaluate the overheads in area and delay as described in the Section \ref{section:\wnameRAM_implementation}.% and update the architecture file.

\begin{table}[t]
  \renewcommand{\arraystretch}{1.1}
   %\vspace{-2mm}
  \centering
  \caption{Properties of the baseline FPGA architecture \\(Intel Arria 10 GX 900 like)}
  \label{table:baseline_fpga_resources}
  %\scriptsize
  \footnotesize
  %\resizebox{0.484\textwidth}{!}{%
  %\vspace{-3mm}
  \begin{tabular}{
  | >{\centering\arraybackslash}m{0.3\columnwidth} 
  | >{\centering\arraybackslash}m{0.15\columnwidth}
  | >{\centering\arraybackslash}m{0.25\columnwidth}
  |}
    \hline
    \rowcolor{LightBlue} \centering
    \textbf{Resource} & \textbf{Count} & \textbf{Percentage Area} \\
    \hline
    Logic Blocks & 33962 & 66  \\ \hline
	DSP Slices & 2423 & 18 \\ \hline
	Block RAMs & 1518 & 15  \\ \hline
	DRAM bandwidth & \multicolumn{2}{c|}{2048 bits/clock} \\ \hline
	Channel width & \multicolumn{2}{c|}{300}  \\ \hline
  \end{tabular}
  %}
\end{table}

\subsection{Benchmarks}

\vspace{-1mm}

\begin{table}[t]
  \renewcommand{\arraystretch}{1.1}
   %\vspace{-2mm}
  \centering
  \caption{List of benchmarks used for evaluation (CB = Compute bound, OMB = On-chip memory-bandwidth bound, DBB = DRAM bandwidth bound)}
  \label{table:benchmarks}
  %\scriptsize
  \footnotesize
  %\resizebox{0.484\textwidth}{!}{%
  %\vspace{-3mm}
  \begin{tabular}{
  | >{\centering\arraybackslash}m{0.27\columnwidth} 
  | >{\centering\arraybackslash}m{0.16\columnwidth}
  | >{\centering\arraybackslash}m{0.11\columnwidth}
  | >{\centering\arraybackslash}m{0.10\columnwidth}
  | >{\centering\arraybackslash}m{0.12\columnwidth}
  |}
    \hline
    \rowcolor{LightBlue} \centering
    \textbf{Benchmark} & \textbf{Domain} & \textbf{Scenario Created} & \textbf{Storage} & \textbf{Precision} \\ \hline
    GEMV  & DL    & CB & DRAM  & 8-bit \\ \hline
    FIR filter & DSP & CB & DRAM  & 16-bit \\ \hline
    Eltwise mult & DL    & DBB & DRAM  & HFP8 \\ \hline
    Bulk bitwise-Search & Databases & OMB & BRAM  & 16-bit \\ \hline
    Bulk bitwise-RAID & Data Center & OMB & BRAM  & 20-bit \\ \hline
    Reduction & DL & OMB & BRAM  & Multiple \\ \hline
  \end{tabular}
  %}
\end{table}

We create Verilog designs for several diverse applications to use as benchmarks (Table \ref{table:benchmarks}). 
%These include Deep Learning (matrix vector multiplication, reduction, elementwise multiplication), signal processing (FIR filter), and bulk-bitwise applications (database search, RAID array data recovery). 
We manually map the applications to \wnameRAMs~and instantiate the \wnameRAM~block in Verilog, which is mapped to \wnameRAMs~in the FPGA architecture provided to VTR. During functional verification, a simulation model of \wnameRAM~is used. 
%We also create performance models for each benchmark using Python to perform design space exploration. 
We create different scenarios (compute bound, DRAM bandwidth bound and on-chip memory bound) in these applications.
%to observe the impact of adding \wnameRAMs on performance. 
%We leave evaluating the benefits of adding \wnameRAMs~for full system applications as future work. We expect the speedups to be similar or higher because the overhead of transposing data (from/to DRAM) will get amortized over multiple kernels.

\textbf{General Matrix Vector Multiplication (GEMV):} 
GEMV is a fundamental operation in DL applications. It is used in CNNs, LSTM cells and in MLPs. We consider GEMV workloads from DeepBench benchmarks \cite{deepbench} - \textit{LSTM,h=512,t=50 }and \textit{GRU,h=512,t=1}. 8-bit integer precision with 27-bit accumulation is used. 
On the baseline FPGA, compute units are implemented using efficient chaining of DSPs. On the proposed FPGA, compute units based on \wnameRAMs~are additionally deployed, because many RAM blocks are available after mapping the baseline design on the proposed FPGA. Efficient OOOR-based dot product algorithm is used.
%used with the vector being outside the RAM. The intelligent dot product algorithm described in Section \ref{section:ooor} is implemented. 
Partial sums are read out from the \wname~blocks and accumulated using a pipelined bit-serial tree \cite{greg_stitt_fpga_bit_serial}. 
No online data transpose is required - the weight matrix is transposed offline and pinned into \wnameRAM~blocks; the vector is streamed and does not need to be transposed because it is outside the RAM.
%This application is very compute intensive and hence is a good candidate of acceleration using \wnameRAMs. 
Since both DSP based and \wname~based compute units are used, reduction in data movement is not expected. 
%The goal is to quantify the throughput improvement of the FPGA by adding \wnameRAMs.

\textbf{Finite Impulse Response (FIR) Filter:} 
FIR filters are a common DSP application. We consider an FIR filter with 128 taps (similar results were observed for 256 taps; not shown here because of space constraints).
%Since only additions are involved, DSP slices are not used in this application. 
Input operands are streamed onto the FPGA through the DRAM interface. The baseline FPGA uses an efficient implementation of FIR filter using systolic DSP chaining \cite{altera_high_perf_filters}. The proposed FPGA uses \wnameRAMs~for computation along with DSP chains. 
Logic blocks were used for control logic.
%We also consider the case where logic blocks were used for computation in both baseline and proposed FPGAs, but .
Operands are transposed on-the-fly and loaded into multiple \wnameRAMs~in parallel.
%Instruction generation control logic is triggered to start sending instructions to the \wnameRAMs. 
While some \wnameRAMs~are computing, other \wnameRAMs~are loaded in a pipelined manner to improve parallelism. When a \wnameRAM~finishes computing, its results are unloaded and sent to DRAM, and the process starts again until all inputs are processed. We call this the Load-Compute-Unload (LCU) pipeline. 
%Although a large amount of data is loaded from DRAM, this application was not DRAM limited . 
%Limited DRAM bandwidth results is large fanout to feed the compute units, which reduces frequency. So, more DRAM bandwidth would have helped the frequency.
In this application, the \wnameRAM-to-\wnameRAM~chaining (Section \ref{section:chaining}) feature is used to share inputs between neighboring blocks.

\textbf{Elementwise multiplication:} Elementwise multiplications are commonly used in DL, for example, in normalization layers and Winograd based convolution layers. 
We consider an application involving elementwise multiplication of two arrays of 100K elements. Floating point data with precision of HFP8 \cite{hfp8} is used. We showcase here that \wnameRAMs~are adaptable to any custom precision. The operands are read from DRAM and the results are written to DRAM. This is a DRAM bandwidth bound application because of low arithmetic intensity. We observed that the number of LBs used (16748) was significantly higher than in the baseline FPGA (649). To saturate the DRAM bandwidth available on the chip, many swizzle logic instances are required. 
%Each swizzle logic requires about 165 LBs. 
To reduce this impact, we plan to explore hardening swizzle logic or adding Transpose Memory Units \cite{neural_cache}.

\textbf{Bulk bitwise operations:} Bulk bitwise operations (like AND, OR, XOR, etc) are commonly used in databases, encryption, DNA sequence alignment, etc. \wnameRAMs~are very efficient at these massively parallel operations because of the presence of mux-based fully configurable PEs.
%We showcase how using \wnameRAMs~can speed up on-chip memory limited applications. 
The operands are assumed to be available in BRAMs in the right layout. The speedup seen in these applications attributed to the effective increase in on-chip memory bandwidth because 160 bits can be operated upon in 1 cycle in a \wnameRAM, compared to only 40 bits from a BRAM in the baseline FPGA. 
We consider two applications in this category.

\textit{Database search:} In this application, records matching a key are searched. If a record matches the key, it is replaced with a special marker data (like constant 0). 
Each operand is bitwise XOR'ed with the key. Bitwise OR reduction is performed on the result. And then a bitwise ANDing operation is performed to zero-out the operands that match the key. 256 RAM blocks are used to store operands. In \wnameRAM, 7 data elements are stored in each column and temporary results consume 16 rows. The key is outside the RAM. 

\textit{RAID data recovery:}
In RAID (Redundancy Array of Independent Disks) arrays, parity protection is used. If a drive in an array fails, the remaining data on other drives is combined with the parity data (using XOR) to reconstruct the missing data. These numerous parallel XOR operations with the parity data can be accelerated using an FPGA. 
%We can use \wnameRAMs~to make such operations efficient. 
%We also demonstrate a different way of using \wnameRAMs~in this study - bit-parallel instead of bit-serial.
Instead of storing operands in a transposed format (bits of one operand in multiple rows), we use an un-transposed data layout where we store bits of one operand in one row and bits of the second operand are in another row.
%The bits are aligned though (e.g. bit 0 of each operand is in one column, bit 1 of each operand is in another column and so on). 
This works for logical operations like bitwise XOR where there is no dependency/communication between consecutive bits, and avoids the overhead of transposing data. 
%Performing an XOR operation between operands stored on two rows takes 1 cycle. 

\textbf{Reduction:}
Reduction (or accumulation) is heavily used in DL and DSP applications. We design this application to create a scenario of an on-chip memory bandwidth limited application. Data is available in transposed format (computed in RAM by a prior kernel for example). The precision is varied from 4-bit to 20-bit (accumulator size = 32-bit). In the baseline, operands stored in BRAMs are read and successively accumulated using a pipelined adder tree (in LBs). On the proposed FPGA, \wnameRAMs~store the operands. The reduction algorithm from \cite{neural_cache} is used to reduce the elements to 40 partial sums (1 partial sum in each column of the RAM). These intermediate results from multiple \wnameRAMs~are then read out and accumulated using a pipelined bit-serial adder \cite{greg_stitt_fpga_bit_serial} to obtain the result. 
A significantly smaller number of LBs (\textasciitilde2x-3.5x) is required on the proposed FPGA.
%In both baseline and proposed cases, we pack the data into BRAMs to ensure maximum utilization. A total of 32 RAMs is used in each case.

%The designs for the baseline FPGA use LBs and DSPs for compute, and BRAMs for storage. The designs for the proposed FPGA use \wnameRAMs~for compute and storage in addition to using the traditional resources. 
%These designs instantiate the \wnameRAM~hard macro, which is mapped to \wnameRAMs~in the FPGA architecture provided to VTR. During functional verification, a simulation model of \wnameRAM~is used. We also create performance models for each benchmark using Python to perform design space exploration. 
%To optimize performance in all benchmarks for the proposed FPGA, we connect \wnameRAM~blocks and the instruction generation logic to a separate clock domain. This approach reduces routing utilization and simplify control logic. The separate clock domain runs at a higher frequency and performs computation significantly faster, and thus achieving higher speed-up.

%For evaluating \wnameRAMs, we design an FPGA architecture containing \wnameRAMs~as described in the Section \ref{section:baseline_vs_proposed_arch} . 

\subsection{Implementation Details}\label{section:\wnameRAM_implementation}

\textbf{Area}: 
%We use COFFE to quantify the area overhead of modifications to a BRAM to convert it into a \wnameRAM. 
Table \ref{table:math_ram_area_breakdown} shows the area breakdown of both architectures of \wnameRAM.
For \wnameD, the area overhead is 1546.78 $um^2$. This represents an increase of 25.4\% in the BRAM tile area compared to the baseline. This overhead is mainly attributed to the addition of 160 PEs and the additional 120 sense amplifiers and write drivers. 
%This increase does not consider modifications to the sequencing logic, because COFFE does not model that logic. However, this overhead will be negligible because it only a few modifications to state machines. 
With BRAMs occupying 15\% of the die size in our baseline FPGA, this overhead corresponds to only 3.8\% increase in the FPGA chip area. 
The overhead for \wnameA~is 493.5 $um^2$. Compared to the baseline, this represents an increase of 8.1\% in BRAM tile area and only 1.2\% increase in FPGA chip area. This overhead is mainly attributed to the addition of 40 PEs.

%\begin{table}[hbt]
%\centering

%\begin{tabular}{ | l|r| }
%\hline
 %   \rowcolor{light-gray}
%	\textbf{Component} & \textbf{\%age} \\ \hline
%	Input and output crossbars & 4.62  \\ \hline
%	Decoders and wordline drivers & 4.86 \\ \hline
%	Write drivers and sense amplifiers & 14.28 \\ \hline
%	Memory cell array & 43.83 \\ \hline
%	Routing (connection and switch blocks) & 21.33 \\ \hline
%	Processing elements & 11.09  \\ \hline
%	\textbf{Total} & \textbf{100.00} \\ \hline
%\end{tabular}

%\caption{Area breakdown of a \wnameRAM~block}
%\label{table:math_ram_area_breakdown}
%\vspace{-4mm}%Put here to reduce too much white space after your table 
%\end{table}

\begin{table}[t]
  \renewcommand{\arraystretch}{1.1}
  % \vspace{-2mm}
  \centering
  \caption{Area breakdown of various RAM blocks}
  \label{table:math_ram_area_breakdown}
  %\scriptsize
  \footnotesize
  %\resizebox{0.484\textwidth}{!}{%
  %\vspace{-3mm}
  \begin{tabular}{
  | >{\centering\arraybackslash}m{0.40\columnwidth} 
  | >{\centering\arraybackslash}m{0.09\columnwidth}
  | >{\centering\arraybackslash}m{0.155\columnwidth}
  | >{\centering\arraybackslash}m{0.155\columnwidth}
  |}
    \hline
    \rowcolor{LightBlue} \centering
    \textbf{Component} & \textbf{BRAM}  & \textbf{\wnameD} &  \textbf{\wnameA} \\
    \hline
   	Input and output crossbars           & 5.6   & 4.5             &   5.2 \\ \hline
	Decoders \& wordline drivers        & 7.8     & 6.3             &   7.3 \\ \hline
	Write drivers \& sense amps.   & 6.9     & 14.0            &   6.4 \\ \hline
	Memory cell array                    & 53.4     & 43.0            &  49.6 \\ \hline
	Routing (conn. \& switch)     & 26.0 & 20.9            &  24.1 \\ \hline
	Processing elements                  & 0    & 11.1            &  7.1 \\ \hline
	\textbf{Total (\%)}                       & \textbf{100}         & \textbf{100} & \textbf{100} \\ \hline
  \end{tabular}
  %}
\end{table}

\textbf{Frequency}:
We use the COFFE to obtain the overhead in frequency of operation of a \wnameRAM~in Hybrid mode, compared to a BRAM (735 MHz). 
For \wnameD, the cycle duration increases to 1.25x (588 MHz). This is mainly attributed to performing read, compute (PE circuitry delay) and write in the same cycle. For \wnameA, the cycle duration increases to 2.5x (294 MHz). This is because 4 reads and 2 writes are done successively as described in Section \ref{processing_element_arch}. 
A lower frequency of the \wnameRAM~is not a concern because realistic FPGA designs typically are constrained by soft-logic and routing delays, so designs do not achieve high frequencies like those of individual BRAMs (735 MHz in this case).
In Memory mode, the delay overhead is negligible because there is only one additional mux in the write path; the read path remains unchanged.

%In a simple implementation, the improved frequency of \wnameRAM~blocks may not provide an overall benefit, because the frequency of operation of the design would likely be governed by soft logic. However, employing double pumping can alleviate this problem \cite{multi_pump_multi_port_rf} \cite{multi_pumping_resource_reduction} \cite{multipumping_dsp_xiling_fpgas}. That is, we operate the RAM blocks (\wnameRAM/CCB, depending on the design) at double the frequency ($2f$) used for the instruction generation logic ($f$). This requires simple time multiplexing logic at the inputs, working at $2f$ and also requires splitting instruction generation logic into two smaller FSMs compared to one larger FSM. 
%We also observed that the benefit of double pumping went away when one instruction generation logic's outputs were fanned out to a large number of RAM blocks. So, the best fanout was experimentally determined for each design. These modifications ensure that \wnameRAM~based designs run at faster frequencies compared to CCB based designs and hence provide more speedup. 

%\textbf{FPGA Configuration Memory}:
%The mode of operation of \wnameRAM~is determined by a configuration SRAM cell, which is configured during programming the FPGA. These additional configuration bits represent a negligible 2.43 Kb increase to the 127 Mb compressed bitstream of our baseline device.

\textbf{Routing}:
The interface of a \wnameRAM~block to the programmable routing is not changed compared to that of a BRAM. The only change is the addition of two pins, which are used for direct connections between neighboring BRAMs. These do not impact the programmable interconnect %(connection boxes, switch boxes, routing channel) 
directly, but do increase the pin density. 
%The final interface of a \wnameRAM~block includes the standard pins for a BRAM - \texttt{clock}, \texttt{address}, \texttt{data\_in}, \texttt{write\_enable}, \texttt{data\_out}, and the two new bidirectional pins - \texttt{pe\_top} and \texttt{pe\_bottom}.

\textbf{CCB}: The implementation of CCB \cite{ccb} is based on a BRAM with 128x128 geometry. 
The area overhead for the CCB block evaluated in \cite{ccb} does not include the area of the additional sense amplifiers and write drivers. 
In our re-implementation of CCB, the total area overhead comes out to be 872.64 $um^2$, which is a 16.8\% increase at the  block level and 2.5\% at the chip level in the Arria-10-like FPGA used in this study.
%We refer to this implementation as CCB-D (delay optimized).
The frequency of operation of the CCB evaluated in \cite{ccb} is 1.6x (469 MHz) compared to the baseline BRAM. Table \ref{table:diff_ccb_comefa} shows the differences between CCB and \wname.

\begin{table}[t]
  \renewcommand{\arraystretch}{1.1}
  \centering
  \caption{Differences between CCB and \wname}
  \label{table:diff_ccb_comefa}
  %\scriptsize
  \scriptsize
  %\resizebox{0.484\textwidth}{!}{%
  %\vspace{-3mm}
  \begin{tabular}{
  | >{\raggedright\arraybackslash}m{0.40\columnwidth} 
  | >{\centering\arraybackslash}m{0.08\columnwidth}
  | >{\centering\arraybackslash}m{0.14\columnwidth}
  | >{\centering\arraybackslash}m{0.14\columnwidth}
  |}
    \hline
    \rowcolor{LightBlue} \centering
    \textbf{Property} & \textbf{CCB} & \textbf{\wname-D} & \textbf{\wname-A}\\
    \hline

Activate two wordlines at the same time on one port    & Yes   & No   & No  \\ \hline
Additional voltage source required     & Yes   & No    & No \\ \hline
Additional row decoder required     & Yes   & No & No\\ \hline
Changes in sense amplifiers    & Yes   & No     & No \\ \hline
Additional sense amplifiers    & Yes  & Yes & No \\ \hline
Sense amp cycling     & No    & No & Yes \\ \hline
Compute uses dual-port behavior     & No    & Yes & Yes \\ \hline
Generic/Flexible PE    & No    & Yes & Yes \\ \hline
Shift between RAM blocks    & No    & Yes & Yes \\ \hline
Floating point support     & No    & Yes & Yes \\ \hline
Flip-flops in PE to store operands    & No    & No & Yes \\ \hline
%Modifications to sequencing logic     & Yes   & Yes & Yes\\ \hline
Parallelism   & 128   & 160 & 160\\ \hline
Application(s) demonstrated  & DL    & Many & Many\\ \hline
%Cycles per operation     & 1     & 1 & 1\\ \hline
Clock duration overhead  & 60\%   & 25\% &125\% \\ \hline
Area overhead (block) & 16.8\%* & 25.4\% &8.1\% \\ \hline
Area overhead (chip)  & 2.5\%* & 3.8\% & 1.2\% \\ \hline
%Targeted for FPGAs & No    & Yes   & Yes & Yes\\ \hline
Column multiplexing & No & No & Yes \\ \hline 
Practicality & Low & Medium & High \\ \hline 

    \hline
  \end{tabular}
  %}
  \scriptsize
  \vspace{-2mm}
    \begin{flushleft}
	\item *includes overhead of additional sense amplifiers and write drivers.
	\end{flushleft}
  %\vspace{-4mm}
  
\end{table}
\vspace{-2mm}
\vspace{1mm}
\section{Results} \label{section:results}
%In this section, we quantitatively analyze the benefits of our proposal. We look at the peak throughput improvement obtained by converting BRAMs into \wnameRAMs~on our baseline FPGA. Then, we look at the speedup obtained for a diverse set of applications.

\vspace{-2mm}
\subsection{Throughput Comparison}
%In this section, we evaluate the peak throughput improvement obtained by enhancing the baseline FPGA with the proposed \wnameRAMs. 
%Peak throughpt.. Put one block and find freq. This is a common method for finidng peak throughut.

To evaluate the peak throughput, we consider the MAC (multiply-accumulate) operation, which is the most common operation in DSP and DL applications. We use common fixed-point precisions - 4-bit (accumulator=16-bit), 8-bit (acc=27-bit) and 16-bit (acc=36-bit), as well as floating-point precisions - HFP8 (\{exp=4, frac=3\} and  acc=\{exp=6, frac=9\}) \cite{hfp8} and IEEE FP16 (acc=IEEE FP32). 
We compare the throughput of \wnameRAMs~to the traditional compute units (LBs and DSPs). For LBs, we synthesize, place and route one MAC onto the FPGA and determine the operating frequency and resource utilization. We then calculate the throughput by optimistically assuming that we can fill the FPGA at the same operating frequency. This serves the purpose for evaluating peak throughput.
%This does not consider the frequency degradation as the FPGA is filled, but serves the purpose for evaluating peak throughput. 
For DSPs, MACs are created and taken through a similar process. 
%Note that for some precisions like fixed-point 8-bit and 16-bit, two MACs are implemented on one Arria-10-like DSP, but for precisions like IEEE FP16, 1 MAC is implemented in the DSP slice. 
The DSPs do not natively support FP16 and HFP8 precisions, so MACs for these precisions are designed using soft logic and DSPs. For \wnameRAMs, 160 MACs are implemented in parallel by instantiating one \wnameRAM~and an instruction generation FSM. 

\begin{figure}[t]
\centering
\includegraphics[width=\linewidth]{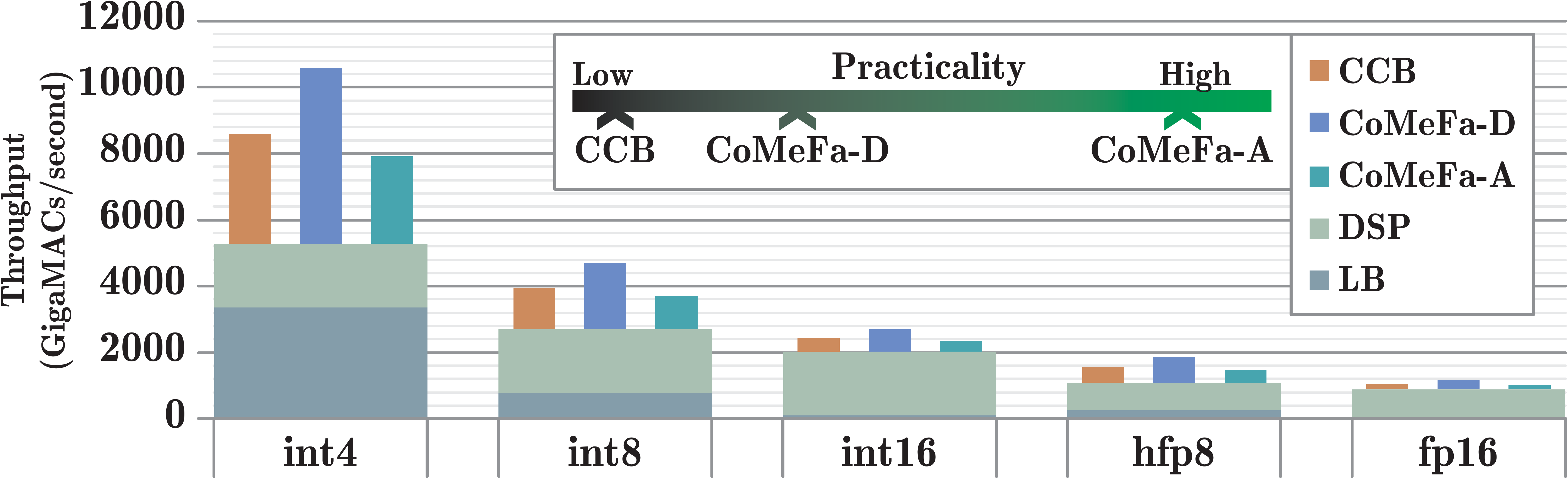}
%\vspace{-4mm}
\caption{Peak throughput for MAC operations for the whole FPGA for various precisions}
%%\wnameRAMs~Provide significant additional throughput in each case.
\label{fig:throughput_comefa}
\vspace{-2mm}
\end{figure}

Figure \ref{fig:throughput_comefa} shows the peak throughput for each precision obtained from each different computing resource in GigaMACs/second. We observe that the throughput of the FPGA increases by 2x, 1.7x, 1.3x, 1.7x and 1.3x for int4, int8, int16, hfp8 and fp16 respectively by adding \wnameD~RAMs. Similarly, the throughput of the FPGA increases by 1.5x, 1.36x, 1.16x, 1.36x and 1.15x for int4, int8, int16, hfp8 and fp16 respectively by adding \wnameA~RAMs. \wnameRAM~throughput reduces as the precision increases, due to the bit-serial nature of computation in \wnameRAMs. 
%Operations in large precisions like int16 consume lot of cycles leading to a smaller throughput enhancement compared to smaller precisions like int4 or int8. 
%\textbf{While \wnameRAM~throughput scales nicely with precision, DSP slices have hard precision support and do not scale well.}
%DSP throughput stays constant in going from int4 to int8 to int16 because of the hardened 18x19 support.
%Among the floating point precisions, we see a similar pattern. 
\wnameRAMs~can be used for computing in any precision, unlike DSPs. The frequency of operation of \wnameRAMs~does not change significantly with changing precision, unlike LBs.

\vspace{-1mm}
\subsection{Speedup and Energy Benefits}

\begin{figure}[t]
\centering
\includegraphics[width=\linewidth]{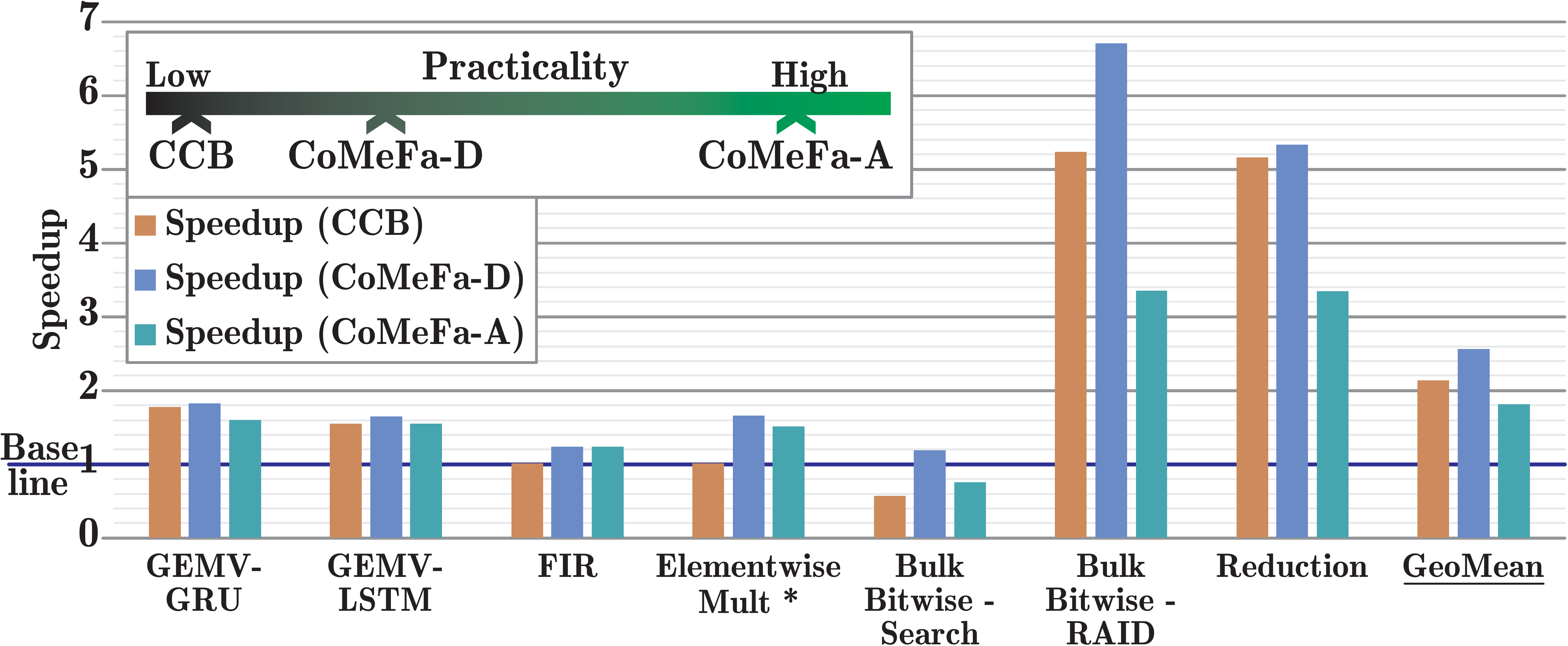}
\vspace{-5mm}
\caption{Speedups obtained for different FPGA architectures for various benchmarks. $*$ implies no DRAM bandwidth limitation.}
\label{fig:all_speedups}
\vspace{-1mm}

\end{figure}

Figure \ref{fig:all_speedups} shows the maximum speedup obtained by using \wnameRAMs~across benchmarks. 
We see significant speedups by using \wnameRAMs~in the compute bound applications because of the augmented compute throughput provided by the FPGA.
For GEMV benchmarks, speedups of upto 81\% are seen in \wname-D~and upto 59\% in \wname-A. With CCB, the max speedup was 72\%. We observed some erraticness in results across seeds because the DSPs were \textasciitilde99\% utilized, so we considered the maximum frequency instead of averaging the frequency across many seeds as we do in other applications.
A speedup of 22\% is seen in the FIR benchmark for both \wname-D~and \wname-A. That's because the frequency of operation of the overall design was \textasciitilde215MHz in both cases.
The FIR benchmark uses chaining of RAMs, which is not supported by CCB. So, no speedup is considered compared to the baseline.

Since the Elementwise multiplication benchmark is limited by DRAM bandwidth, no speedup is seen by using \wnameRAMs. %This is expected as 
\wnameRAMs~are targeted to improve the compute throughput of the FPGA, not the DRAM bandwidth. %Rather change it to - "not compensate for DRAM bandwidth limitations" at the end 
%DRAM bandwidth on FPGAs has been steadily increasing because of demands of modern applications. 
If we remove the restriction of DRAM bandwidth and assume that all the compute units (\wnameRAMs~as well as DSPs/LBs) can be fed with data, then speedups of 65\% and 50\% can be seen on \wname-D and \wname-A FPGAs respectively.
Since CCB does not support floating-point operations, the speedup for this benchmark for CCB is shown as 0\%.

The Search benchmark is sped up by 18\% for \wnameD. The design on baseline FPGA had the highest frequency of operation because of very simple operations done in soft logic. No speedup is seen using \wnameA~RAMs because of the low frequency of operation.
This application is not sped up by using CCB either. CCB takes \textasciitilde2x cycles compared to \wnameRAM~because of the inflexibility of the processing elements that only support a few operations. E.g. AND operation can be done in 2 cycles in CCB, compared to 1 cycle in \wnameRAM.
The RAID application is sped up by 6.7x in \wnameD, 3.35x in \wnameA and 5.2x in CCB. The baseline frequencies were very high in this case also, but the difference in number of cycles enabled the significant speedups.
The speedups for the Reduction benchmark (4-bit precision) were 5.3x in \wnameD, 3.3x in \wnameA~and 5.1x in CCB. 
%A significant reduction in LB usage (\textasciitilde2x-3.5x) is also seen in Search, RAID and Reduction benchmarks.

\begin{figure}[t]
\centering
\includegraphics[width=\linewidth]{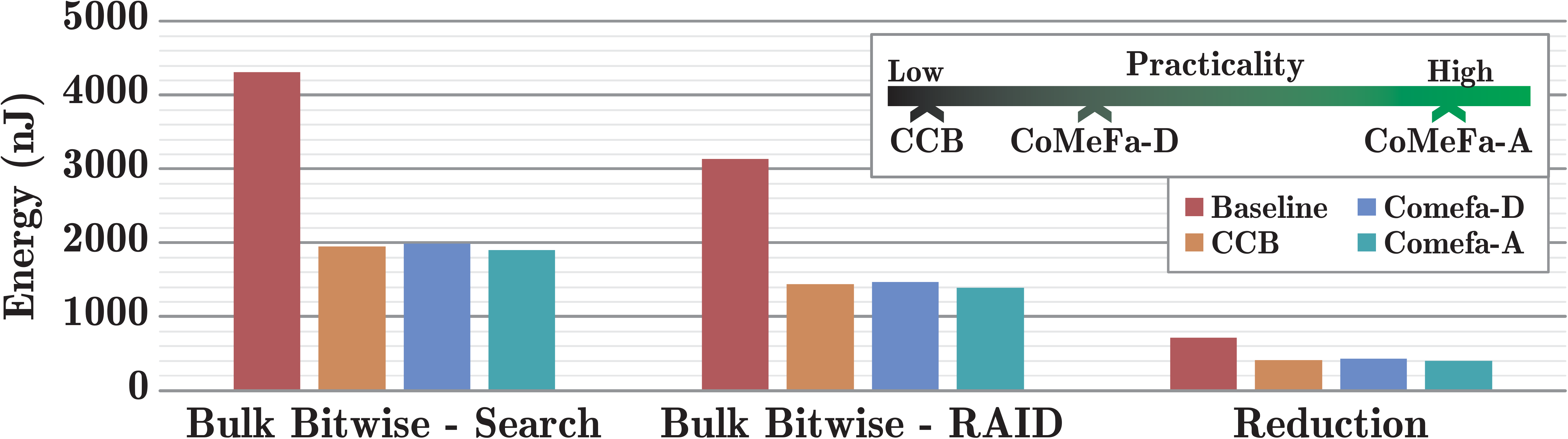}
\vspace{-2mm}
\caption{Energy savings by reduced data movement in on-chip memory bandwidth limited benchmarks}
\label{fig:energy}
\vspace{-3mm}

\end{figure}

In the on-chip memory bandwidth bound benchmarks (Search, RAID, Reduction), upto 62\% LBs are used in \wname ~compared to baseline. That is because no external LUTs are needed when \wnameRAMs~are used. Routing WL reduction of upto 68\% is seen, which directly correlates to reduction in data movement. Results from our energy model are shown in Figure \ref{fig:energy}. We see an energy reduction of upto 56\% in \wnameA~and upto 52\% in \wnameD.

\vspace{-2mm}
\subsection{Application Co-mapping}
\begin{figure}[t]
\centering
\includegraphics[width=0.95\linewidth]{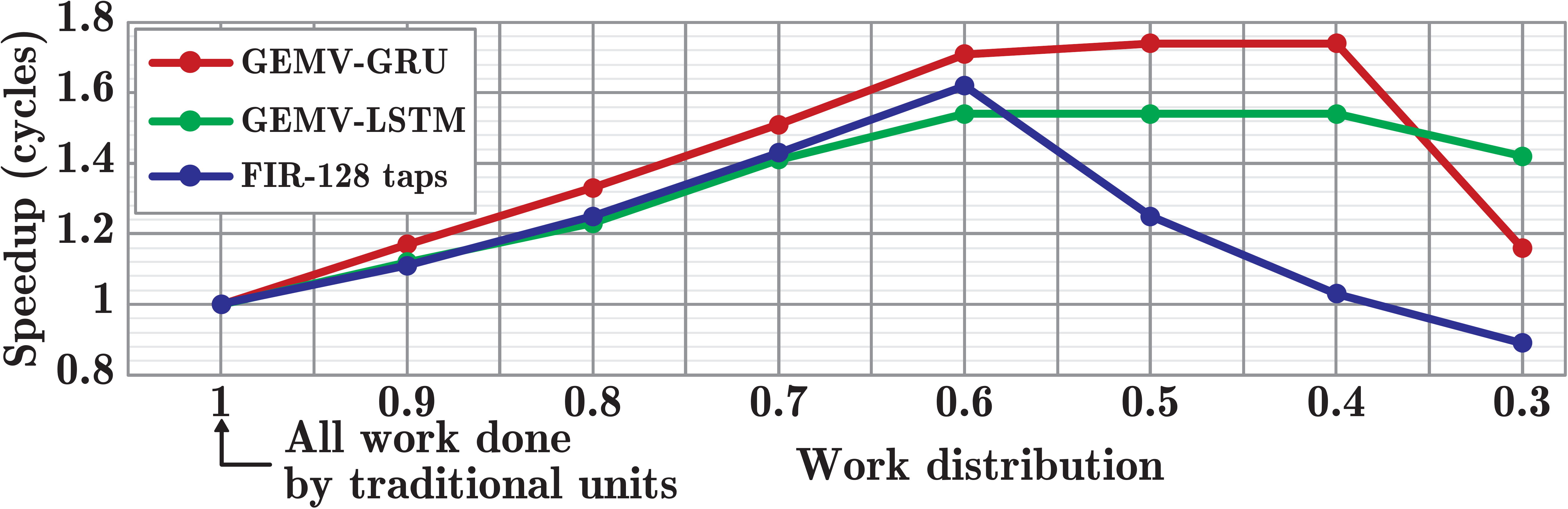}
%\vspace{-5mm}
%\caption{Variation of speed-up (in cycles) for different GEMV and FIR applications with the amount of work distributed between DSPs and \wnameRAMs}
\caption{Illustration of increase in speed-up (based on cycles) by partitioning the application between DSPs and \wnameRAMs.}
%The increased compute capability of \wnameRAMs ~can provide up to 70\% improvement in the enhanced FPGA by partitioning the application between DSPs and \wnameRAMs. 
\label{fig:speedup_with_work_distribution}
\end{figure}

%\wnameRAMs~do not  compete with DSPs and LBs, but enhance the FPGA's compute throughput. 
\wnameRAMs~supplement DSPs and LBs as compute units, and enhance the FPGA's compute throughput.
Appropriately dividing and mapping parts of an application to \wnameRAMs ~and traditional compute units is key. For GEMV and FIR applications, we analytically explore the effect of varying work distribution between \wnameRAMs~and DSPs/LBs on the proposed FPGA. The results are shown in Figure \ref{fig:speedup_with_work_distribution}. 
We see that as more work is given to \wnameRAMs, more speedup can be obtained upto a limit, after which the overheads (loading, unloading, serial compute) associated with \wnameRAMs~can start dominating and reduce the overall speedup.
This sweet spot is different for each application.

%\begin{figure}
%\begin{subfigure}{.25\textwidth}
%  \centering
%  \includegraphics[width=.9\linewidth]{figures/mvm_speedup_vs_work_dist.png}
%  %\caption{1b}
% \label{fig:mvm_work_dist_vs_speedup}
%\end{subfigure}
%\begin{subfigure}{.25\textwidth}
%  \centering
%  \includegraphics[width=.9\linewidth]{figures/mvm_speedup_vs_precision.png}
%  %\caption{1a}
%  \label{fig:mvm_precision_vs_speedup}
%\end{subfigure}%
%\caption{Graphs showing the variation of speedup for MVM with (a) work distribution (for int8 precision), and (b) %precision and bit-slicing}
%\label{fig:mvm_charts}
%\end{figure}

\subsection{Adaptability to Precision}
\begin{figure}[t]
\centering
\includegraphics[width=\linewidth]{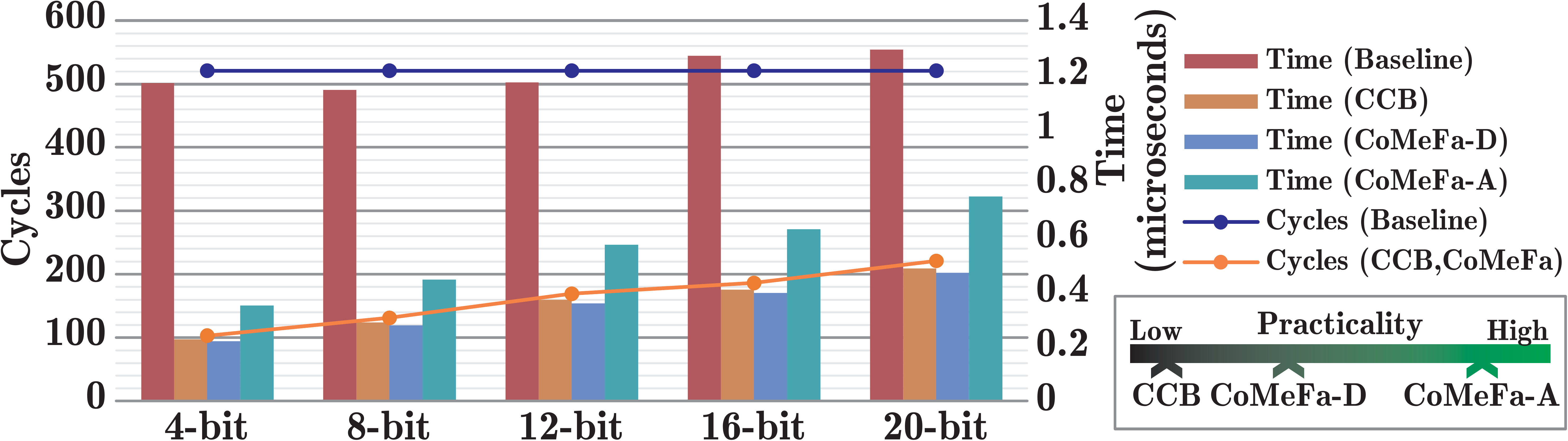}
\vspace{-5mm}
\caption{Sweeping precision in the Reduction benchmark}
\label{fig:accum_precision_sweep}
\vspace{-2mm}
\end{figure}

\wnameRAMs~can be used for efficiently computing in any custom precision. 
Figure \ref{fig:accum_precision_sweep} shows the results of sweeping the precision from 4-bit to 20-bit in the Reduction benchmark. We see  speedups ranging from 5.3x (3.3x) to 2.7x (1.7x) with \wnameD~(\wnameA) as precision increases. \wnameD~is 3\% better than CCB owing to the improved frequency achieved by the design. The baseline takes the same number of cycles for each precision because of the bit-parallel nature of compute. But the number of cycles taken increases as the precision increases when \wnameRAMs~are used. This is because of bit-serial arithmetic and illustrates that applications using smaller precisions are better suited for \wnameRAMs.
Note that the frequency of operation stays constant for \wnameRAMs~because the hardware architecture stays the same. For the baseline, the frequency decreases slightly as the precision increases.  

%Figure \ref{fig:all_speedups} shows the speedup obtained for this application for 20-bit precision. The frequency of operation was 327 MHz in the baseline, 478 MHz for \wnameD, 294 MHz for \wnameA and 453 MHz for CCB. The baseline used 2.7x more LBs compared to when \wnameRAMs were used.

\vspace{-1mm}
\section{Conclusion} \label{section:conclusion}

In this paper, we propose augmenting the compute density of FPGAs by modifying BRAMs into new blocks called \wnameRAMs, which are ideal for enhancing applications with inherent parallelism like deep learning and signal processing.
%BRAMs support true dual ported operation and hence can read two operands in one cycle. We exploit this behavior and add processing elements near the sense amplifiers. 
To the best of our knowledge, this is the first work that (1) utilizes the dual-port nature of BRAMs to achieve in-BRAM compute, (2) deploys configurable 1-bit processing elements inside an FPGA BRAM, and (3) applies in-BRAM compute to DL and non-DL applications on FPGAs.
%We propose two different architectures of \wnameRAMs, \wnameD~and \wnameA, demonstrating a tradeoff between area, delay and practicality. Using intelligent algorithms and efficient mapping of applications to an FPGA containing \wnameRAMs, we observe an average speedup of 2.9x and 2x using \wnameD~and \wnameA~respectively, across several representative benchmarks. 
%We observe that the area overhead of making these enhancements to BRAMs is \textasciitilde12.47\% at the block level, which translates to \textasciitilde2.74\% increase in the FPGA die area for our baseline Stratix 10-like FPGA. A frequency reduction of \textasciitilde25\% is observed in the Hybrid Mode, compared to \textasciitilde60\% reduction in state-of-the-art Compute Capable BRAMs. The frequency remains the same as a regular BRAM in memory mode. %The programmable routing fabric of the FPGA is not impacted by adding \wnameRAMs.
%We see a peak throughput improvement ranging from 2.7x to 1.5x for various precisions for MAC operations, compared to a Stratix 10-like baseline FPGA. 
%\wnameRAMs~can be deployed for a variety of applications like DL, DSP, parity/encoding, image processing, databases, etc. 
%For a common DSP application (moving average filter) using 16-bit fixed-point precision, we see a speedup of up to X\%. Similarly, for a Deep Learning application (GEMV), we see speedupds ranging from A\% to B\% depending upon C.
%\wnameRAMs~bring compute closer to memory on an FPGA. 
With improvements to compute density and reduction in power consumption, we believe that converting some or all BRAMs on FPGAs to \wnameRAMs~can be a significant step towards closing the performance gap between FPGAs and ASICs.

%This work as shown encouraging results 

%\input{acknowledgment}
%\balance
\bibliographystyle{IEEEtran}
\bibliography{bibliography}

\begin{thebibliography}{10}
\providecommand{\url}[1]{#1}
\csname url@samestyle\endcsname
\providecommand{\newblock}{\relax}
\providecommand{\bibinfo}[2]{#2}
\providecommand{\BIBentrySTDinterwordspacing}{\spaceskip=0pt\relax}
\providecommand{\BIBentryALTinterwordstretchfactor}{4}
\providecommand{\BIBentryALTinterwordspacing}{\spaceskip=\fontdimen2\font plus
\BIBentryALTinterwordstretchfactor\fontdimen3\font minus
  \fontdimen4\font\relax}
\providecommand{\BIBforeignlanguage}[2]{{%
\expandafter\ifx\csname l@#1\endcsname\relax
\typeout{** WARNING: IEEEtran.bst: No hyphenation pattern has been}%
\typeout{** loaded for the language `#1'. Using the pattern for}%
\typeout{** the default language instead.}%
\else
\language=\csname l@#1\endcsname
\fi
#2}}
\providecommand{\BIBdecl}{\relax}
\BIBdecl

\bibitem{xilinx_ultrascale_bram}
\BIBentryALTinterwordspacing
Xilinx. (2021) {UltraScale Architecture Memory Resources}. [Online]. Available:
  \url{https://www.xilinx.com/support/documentation/user_guides/ug573-ultrascale-memory-resources.pdf}
\BIBentrySTDinterwordspacing

\bibitem{arria10_overview}
\BIBentryALTinterwordspacing
Intel, ``{Intel Arria 10 Device Overview},'' 2021. [Online]. Available:
  \url{https://www.intel.com/content/www/us/en/docs/programmable/683332/current/device-overview.html}
\BIBentrySTDinterwordspacing

\bibitem{neural_cache}
\BIBentryALTinterwordspacing
C.~Eckert \emph{et~al.}, ``{Neural Cache: Bit-Serial in-Cache Acceleration of
  Deep Neural Networks},'' in \emph{Proceedings of the 45th Annual
  International Symposium on Computer Architecture}, ser. ISCA '18.\hskip 1em
  plus 0.5em minus 0.4em\relax IEEE Press, 2018, p. 383–396. [Online].
  Available: \url{https://doi.org/10.1109/ISCA.2018.00040}
\BIBentrySTDinterwordspacing

\bibitem{greg_stitt_fpga_bit_serial}
\BIBentryALTinterwordspacing
A.~Landy and G.~Stitt, ``{Serial Arithmetic Strategies for Improving FPGA
  Throughput},'' \emph{ACM Trans. Embed. Comput. Syst.}, vol.~16, no.~3, jul
  2017. [Online]. Available: \url{https://doi.org/10.1145/2996459}
\BIBentrySTDinterwordspacing

\bibitem{fpga_bit_serial_2}
------, ``{Revisiting Serial Arithmetic: A Performance and Tradeoff Analysis
  for Parallel Applications on Modern FPGAs},'' in \emph{2015 IEEE 23rd Annual
  International Symposium on Field-Programmable Custom Computing Machines},
  2015, pp. 9--16.

\bibitem{pim_workload_perspective}
S.~Ghose, A.~Boroumand, J.~S. Kim, J.~Gómez-Luna, and O.~Mutlu,
  ``{Processing-In-Memory: A Workload-Driven Perspective},'' \emph{IBM Journal
  of Research and Development}, vol.~63, no.~6, pp. 3:1--3:19, 2019.

\bibitem{isaac}
A.~{Shafiee} \emph{et~al.}, ``{ISAAC: A Convolutional Neural Network
  Accelerator with In-Situ Analog Arithmetic in Crossbars},'' in \emph{2016
  ACM/IEEE 43rd Annual International Symposium on Computer Architecture
  (ISCA)}, 2016, pp. 14--26.

\bibitem{prime}
P.~{Chi} \emph{et~al.}, ``{PRIME: A Novel Processing-in-Memory Architecture for
  Neural Network Computation in ReRAM-Based Main Memory},'' in \emph{2016
  ACM/IEEE 43rd Annual International Symposium on Computer Architecture
  (ISCA)}, 2016, pp. 27--39.

\bibitem{floatpim}
M.~Imani, S.~Gupta, Y.~Kim, and T.~Rosing, ``{FloatPIM: In-Memory Acceleration
  of Deep Neural Network Training with High Precision},'' in \emph{Proceedings
  of the 46th International Symposium on Computer Architecture}, 2019, p.
  802–815.

\bibitem{drisa}
S.~Li, D.~Niu, K.~T. Malladi, H.~Zheng, B.~Brennan, and Y.~Xie, ``{DRISA: A
  DRAM-based Reconfigurable In-Situ Accelerator},'' in \emph{2017 50th Annual
  IEEE/ACM International Symposium on Microarchitecture (MICRO)}, 2017, pp.
  288--301.

\bibitem{ambit}
V.~Seshadri, D.~Lee, T.~Mullins, H.~Hassan, A.~Boroumand, J.~Kim, M.~A. Kozuch,
  O.~Mutlu, P.~B. Gibbons, and T.~C. Mowry, ``{Ambit: In-Memory Accelerator for
  Bulk Bitwise Operations Using Commodity DRAM Technology},'' in \emph{2017
  50th Annual IEEE/ACM International Symposium on Microarchitecture (MICRO)},
  2017, pp. 273--287.

\bibitem{ComputeDRAM}
\BIBentryALTinterwordspacing
F.~Gao, G.~Tziantzioulis, and D.~Wentzlaff, ``{ComputeDRAM: In-Memory Compute
  Using Off-the-Shelf DRAMs},'' in \emph{Proceedings of the 52nd Annual
  IEEE/ACM International Symposium on Microarchitecture}, ser. MICRO '52.\hskip
  1em plus 0.5em minus 0.4em\relax New York, NY, USA: Association for Computing
  Machinery, 2019, p. 100–113. [Online]. Available:
  \url{https://doi-org.ezproxy.lib.utexas.edu/10.1145/3352460.3358260}
\BIBentrySTDinterwordspacing

\bibitem{compute_memory_uiuc}
M.~Kang, M.-S. Keel, N.~R. Shanbhag, S.~Eilert, and K.~Curewitz, ``{An
  energy-efficient VLSI architecture for pattern recognition via deep embedding
  of computation in SRAM},'' in \emph{2014 IEEE International Conference on
  Acoustics, Speech and Signal Processing (ICASSP)}, 2014, pp. 8326--8330.

\bibitem{x_sram}
A.~Agrawal, A.~Jaiswal, C.~Lee, and K.~Roy, ``{X-SRAM: Enabling In-Memory
  Boolean Computations in CMOS Static Random Access Memories},'' \emph{IEEE
  Transactions on Circuits and Systems I: Regular Papers}, vol.~65, no.~12, pp.
  4219--4232, 2018.

\bibitem{dima}
M.~Kang, S.~K. Gonugondla, and N.~R. Shanbhag, ``{Deep In-Memory Architectures
  in SRAM: An Analog Approach to Approximate Computing},'' \emph{Proceedings of
  the IEEE}, vol. 108, no.~12, pp. 2251--2275, 2020.

\bibitem{compute_sram}
J.~Wang, X.~Wang, C.~Eckert, A.~Subramaniyan, R.~Das, D.~Blaauw, and
  D.~Sylvester, ``{A 28-nm Compute SRAM With Bit-Serial Logic/Arithmetic
  Operations for Programmable In-Memory Vector Computing},'' \emph{IEEE Journal
  of Solid-State Circuits}, vol.~55, no.~1, pp. 76--86, 2020.

\bibitem{computational_ram}
D.~Elliott, M.~Stumm, W.~Snelgrove, C.~Cojocaru, and R.~Mckenzie,
  ``{Computational RAM: implementing processors in memory},'' \emph{IEEE Design
  Test of Computers}, vol.~16, no.~1, pp. 32--41, 1999.

\bibitem{supreet_logic_in_memory}
S.~Jeloka \emph{et~al.}, ``{A 28 nm Configurable Memory (TCAM/BCAM/SRAM) Using
  Push-Rule 6T Bit Cell Enabling Logic-in-Memory},'' \emph{IEEE Journal of
  Solid-State Circuits}, vol.~51, no.~4, pp. 1009--1021, 2016.

\bibitem{ccb}
X.~Wang, V.~Goyal, J.~Yu, V.~Bertacco, A.~Boutros, E.~Nurvitadhi, C.~Augustine,
  R.~Iyer, and R.~Das, ``{Compute-Capable Block RAMs for Efficient Deep
  Learning Acceleration on FPGAs},'' in \emph{2021 IEEE 29th Annual
  International Symposium on Field-Programmable Custom Computing Machines
  (FCCM)}, 2021, pp. 88--96.

\bibitem{arch_enhancements_in_stratix_v}
\BIBentryALTinterwordspacing
D.~Lewis, D.~Cashman, M.~Chan, J.~Chromczak, G.~Lai, A.~Lee, T.~Vanderhoek, and
  H.~Yu, ``{Architectural Enhancements in Stratix V},'' in \emph{Proceedings of
  the ACM/SIGDA International Symposium on Field Programmable Gate Arrays},
  ser. FPGA '13.\hskip 1em plus 0.5em minus 0.4em\relax New York, NY, USA:
  Association for Computing Machinery, 2013, p. 147–156. [Online]. Available:
  \url{https://doi.org/10.1145/2435264.2435292}
\BIBentrySTDinterwordspacing

\bibitem{arria_10_arch}
J.~Tyhach \emph{et~al.}, ``{Arria 10 Device Architecture},'' in \emph{2015 IEEE
  Custom Integrated Circuits Conference (CICC)}, 2015, pp. 1--8.

\bibitem{dont_forget_the_memory}
\BIBentryALTinterwordspacing
S.~Yazdanshenas, K.~Tatsumura, and V.~Betz, ``{Don't Forget the Memory:
  Automatic Block RAM Modelling, Optimization, and Architecture Exploration},''
  in \emph{Proceedings of the 2017 ACM/SIGDA International Symposium on
  Field-Programmable Gate Arrays}, ser. FPGA '17.\hskip 1em plus 0.5em minus
  0.4em\relax New York, NY, USA: Association for Computing Machinery, 2017, p.
  115–124. [Online]. Available: \url{https://doi.org/10.1145/3020078.3021731}
\BIBentrySTDinterwordspacing

\bibitem{compute_cache}
S.~Aga, S.~Jeloka, A.~Subramaniyan, S.~Narayanasamy, D.~Blaauw, and R.~Das,
  ``{Compute Caches},'' in \emph{2017 IEEE International Symposium on High
  Performance Computer Architecture (HPCA)}, 2017, pp. 481--492.

\bibitem{cache_automaton}
\BIBentryALTinterwordspacing
A.~Subramaniyan, J.~Wang, E.~R.~M. Balasubramanian, D.~Blaauw, D.~Sylvester,
  and R.~Das, ``{Cache Automaton},'' in \emph{Proceedings of the 50th Annual
  IEEE/ACM International Symposium on Microarchitecture}, ser. MICRO-50
  '17.\hskip 1em plus 0.5em minus 0.4em\relax New York, NY, USA: Association
  for Computing Machinery, 2017, p. 259–272. [Online]. Available:
  \url{https://doi.org/10.1145/3123939.3123986}
\BIBentrySTDinterwordspacing

\bibitem{vtr8}
K.~E. Murray, O.~Petelin, S.~Zhong, J.~M. Wang, M.~ElDafrawy, J.-P. Legault,
  E.~Sha, A.~G. Graham, J.~Wu, M.~J.~P. Walker, H.~Zeng, P.~Patros, J.~Luu,
  K.~B. Kent, and V.~Betz, ``{VTR 8: High Performance CAD and Customizable FPGA
  Architecture Modelling},'' \emph{ACM Trans. Reconfigurable Technol. Syst.},
  2020.

\bibitem{coffe2}
S.~Yazdanshenas and V.~Betz, ``{COFFE2: Automatic Modelling and Optimization of
  Complex and Heterogeneous FPGA Architectures},'' \emph{ACM Transactions on
  Reconfigurable Technology and Systems (TRETS)}, vol.~12, no.~1, pp.
  3:1--3:27, January 2019.

\bibitem{asu_ptm}
\BIBentryALTinterwordspacing
A.~S. University. (2012) {Predictive Technology Model}. [Online]. Available:
  \url{http://ptm.asu.edu/}
\BIBentrySTDinterwordspacing

\bibitem{ncsu_freepdk45}
\BIBentryALTinterwordspacing
NCSU. (2018) {FreePDK45}. [Online]. Available:
  \url{https://www.eda.ncsu.edu/wiki/FreePDK45:Contents}
\BIBentrySTDinterwordspacing

\bibitem{gpus_parallel_computing}
S.~W. Keckler, W.~J. Dally, B.~Khailany, M.~Garland, and D.~Glasco, ``{GPUs and
  the Future of Parallel Computing},'' \emph{IEEE Micro}, vol.~31, no.~5, pp.
  7--17, 2011.

\bibitem{Stillmaker201774}
A.~Stillmaker and B.~Baas, ``{Scaling equations for the accurate prediction of
  {CMOS} device performance from 180 nm to 7 nm},'' \emph{Integration, the
  {VLSI} Journal}, vol.~58, pp. 74--81, 2017,
  \url{http://vcl.ece.ucdavis.edu/pubs/2017.02.VLSIintegration.TechScale/}.

\bibitem{arria10_product_table}
\BIBentryALTinterwordspacing
Intel, ``{Intel Arria 10 Product Table},'' 2021. [Online]. Available:
  \url{https://www.intel.cn/content/dam/www/programmable/us/en/pdfs/literature/pt/arria-10-product-table.pdf}
\BIBentrySTDinterwordspacing

\bibitem{arria10_phy_ug}
\BIBentryALTinterwordspacing
------, ``{Intel Arria 10 Transceiver PHY User Guide},'' 2021. [Online].
  Available:
  \url{https://www.intel.cn/content/dam/www/programmable/us/en/pdfs/literature/hb/arria-10/ug_arria10_xcvr_phy.pdf}
\BIBentrySTDinterwordspacing

\bibitem{hmc_user_guide_arria10}
\BIBentryALTinterwordspacing
------, ``{Hybrid Memory Cube Controller IP Core User Guide v16.0},'' 2016.
  [Online]. Available:
  \url{https://www.intel.com/content/www/us/en/docs/programmable/683854/16-0/introduction.html}
\BIBentrySTDinterwordspacing

\bibitem{koios}
A.~Arora, A.~Boutros, D.~Rauch, A.~Rajen, A.~Borda, S.~A. Damghani, S.~Mehta,
  S.~Kate, P.~Patel, K.~B. Kent, V.~Betz, and L.~K. John, ``{Koios: {A} Deep
  Learning Benchmark Suite for {FPGA} Architecture and CAD Research},'' in
  \emph{2021 31st International Conference on Field-Programmable Logic and
  Applications (FPL)}, 2021.

\bibitem{arria10_datasheet}
\BIBentryALTinterwordspacing
Intel, ``{Intel Arria 10 Device Datasheet},'' 2020. [Online]. Available:
  \url{https://www.intel.com.tw/content/dam/www/programmable/us/en/pdfs/literature/hb/arria-10/a10_datasheet.pdf}
\BIBentrySTDinterwordspacing

\bibitem{deepbench}
\BIBentryALTinterwordspacing
S.~Narang. (2016) Baidu deepbench. [Online]. Available:
  \url{https://svail.github.io/DeepBench/}
\BIBentrySTDinterwordspacing

\bibitem{altera_high_perf_filters}
\BIBentryALTinterwordspacing
Altera, ``{Designing Filters for High Performance},'' 2015. [Online].
  Available:
  \url{https://www.intel.cn/content/dam/www/programmable/us/en/pdfs/literature/wp/wp-01260-stratix10-designing-filters-for-high-performance.pdf}
\BIBentrySTDinterwordspacing

\bibitem{hfp8}
\BIBentryALTinterwordspacing
X.~Sun \emph{et~al.}, ``{Hybrid 8-bit Floating Point (HFP8) Training and
  Inference for Deep Neural Networks},'' in \emph{Advances in Neural
  Information Processing Systems}, vol.~32.\hskip 1em plus 0.5em minus
  0.4em\relax Curran Associates, Inc., 2019. [Online]. Available:
  \url{https://proceedings.neurips.cc/paper/2019/file/65fc9fb4897a89789352e211ca2d398f-Paper.pdf}
\BIBentrySTDinterwordspacing

\end{thebibliography}

\end{document}